\documentclass[12pt]{iopart}
\usepackage{iopams}       
\usepackage{graphicx}
\usepackage{amssymb}
\usepackage{amsfonts}
\usepackage{setstack}
\newcommand{\be}{\begin{equation}}
\newcommand{\ee}{\end{equation}}
\newcommand{\ba}{\begin{eqnarray}}
\newcommand{\ea}{\end{eqnarray}}
\newcommand{\half}{{\textstyle \frac{1}{2}}}
\newcommand{\halfs}{{\scriptstyle \frac{1}{2}}}
\renewcommand{\theequation}{\arabic{equation}}

\newcommand{\bn}{{\bar n}}
\newcommand{\bp}{{\bar p}}
\newcommand{\hJ}{{\hat J}}
\newcommand{\bA}{\mathbf{A}}
\newcommand{\bB}{\mathbf{B}}
\newcommand{\bC}{\mathbf{C}}
\newcommand{\bG}{\mathbf{G}}
\newcommand{\bJ}{\mathbf{J}}
\newcommand{\bcP}{{\pmb{\mathcal{P}}}}
\newcommand{\bT}{\mathbf{T}}
\newcommand{\bX}{\mathbf{X}}
\newcommand{\bY}{\mathbf{Y}}

\newcommand{\rd}{\mathrm{d}}
\newcommand{\re}{\mathrm{e}}
\newcommand{\ri}{\mathrm{i}}

\begin{document}
\title[ Correlation in Ising models with holes]%
{Spin-spin correlations in central rows of Ising models with holes}

\author{Helen Au-Yang and Jacques H.H. Perk}
\address{Department of Physics, Oklahoma State University, 
145 Physical Sciences, Stillwater, OK 74078-3072, USA}
\ead{helenperk@yahoo.com, perk@okstate.edu}

\begin{abstract}
In our previous works on infinite horizontal Ising strips of width $m$ alternating with layers
of strings of Ising chains of length $n$, we found the surprising result that the specific heats
are not much different for different values of $N$, the separation of the strings.
For this reason, we study here for $N=1$ the spin-spin correlation in the central row of each strip,
and also the central row of a strings layer.
We show that these can be written as a Toeplitz determinants.
Their generating functions are ratios of two polynomials, which in the limit of infinite vertical size
become square roots of polynomials whose degrees are $m+1$ where $m$ is the size of the strips.
We find the asymptotic behaviors near the critical temperature to be two-dimensional Ising-like.
But in regions not very close to criticality the behavior may be different for
different $m$ and $n$.
Finally, in the appendix we shall present results for generating functions in more general models.
\end{abstract}
\maketitle

\section {Introduction}

There are not many exact results on dimensional and crossover effects in weakly coupled
periodic arrays of boxes or layers in which the interactions are much more pronounced.
Therefore in \cite{HJPih} we introduced a special Ising model in which a large sequence
of identical strips is coupled by sequences of Ising chains. We presented several results
for specific heats, local magnetizations and pair correlations in such ``Ising models with holes,"
without giving their derivations. As these derivations use very different methods, it may be
less confusing to present them separately.

In \cite{HJPihd} we used the dimer method to calculate the free energy and thus also the
specific heat. In this paper we shall use the Clifford algebra approach introduced by
Kaufman \cite{Kaufman} to derive the results for magnetizations and pair correlations
presented in \cite{HJPih}. Besides, several further results are presented in the appendix
for more general layered models.

\begin{figure}[hbt]
\begin{center}
\includegraphics[width=0.75\hsize]{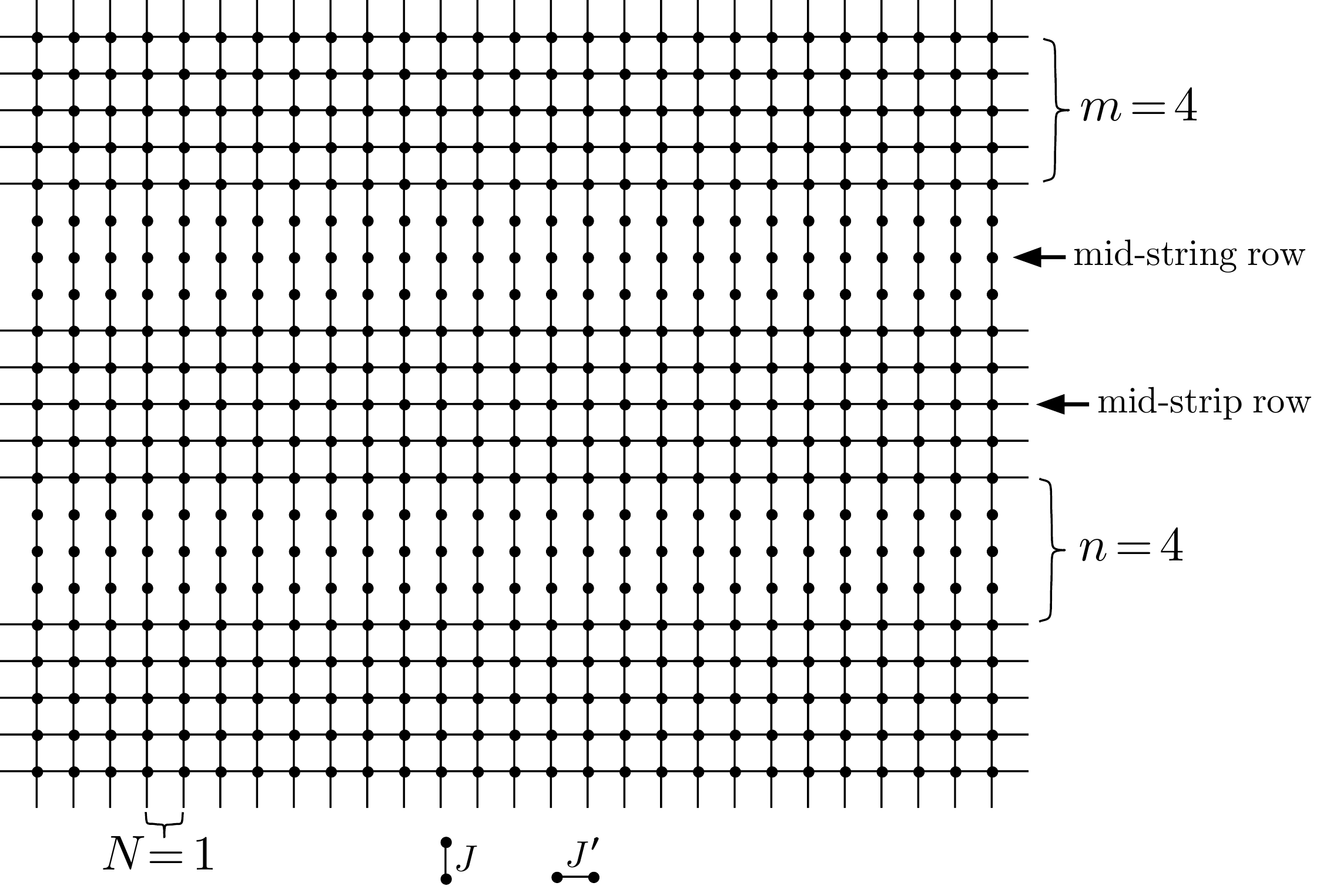}
\caption{Part of special layered Ising model for case with strip width $m=4$
and string length $n=4$. The full model has horizontal size $\bar p$ and
vertical size $p(m+n)$. Pair correlation is calculated
both for mid-strip rows and mid-string rows.}
\end{center}
\label{fig1}
\end{figure}

To be specific, we consider the Ising model consisting of a periodic array of $p$ horizontal
strips of width $m$ and length $\bar p$, which are connected by $\bar p$ vertical strings of length $n$,
see figure \ref{fig1}, which is also  figure 1 of our previous paper \cite{HJPih} with $N=1$.
Again as in \cite{HJPihd}, the horizontal couplings between the nearest-neighbor spins are $J'$,
different from the vertical couplings $J$. We shall require the thermodynamic limit with
$\bar p\to\infty$, but we may leave the vertical period $p$ finite. Then we can derive Toeplitz
determinant formulae for pair correlations in horizontal rows, about which the model is
reflection invariant, that is mid-strip and mid-string rows as indicated in \ref{fig1}.  Since there
are $m+1$ horizontal rows in a strip, to have a central row, we must choose $m$ even, namely
$m=2j$. Similarly, for mid-string rows $n$ must be even.

\section {Clifford algebra approach}

We shall use the Gamma-matrices approach of Kauffman and Onsager \cite{Kaufman,KaufmanO},
which is presented in detail from page 372 on in \cite{KHuang}. However, we use different but
equivalent notations, namely $\mathbf{s}=Z=\sigma^z$, $\mathbf{C}=X=\sigma^x$. We define
the Jordan--Wigner transform
\ba 
\Gamma_{2j-1}=\sigma_1^x\cdots\sigma_{j-1}^x\sigma_{j}^z,\quad
\Gamma_{2j}=-\ri\Gamma_{2j-1}\sigma_{j}^x=\sigma_1^x\cdots\sigma_{j-1}^x\sigma_{j}^y, 
\label{gamma}\ea
for $1\le j\le{\bar p}$ with the periodic boundary condition ${\bar p}+j\equiv j$. The above
equations are identical to (15.43) on page 374 of \cite{KHuang} and these matrices satisfy the anticommutation rule
\be
\Gamma_{j}\Gamma_{k}+\Gamma_{k}\Gamma_{j}=2\delta_{j,k},\quad 1\le j,k\le 2\bar p.
\label{GG}\ee
From (\ref{gamma}), we have
\ba 
\sigma_{j}^z\sigma_{j+1}^z=\ri\Gamma_{2j}\Gamma_{2j+1},\quad
\sigma_{j}^x=\ri\Gamma_{2j-1}\Gamma_{2j}.
\label{sigmazx}\ea
Consequently, the $m+1$ transfer matrices associated with horizontal interactions between
the nearest-neighbor spins pairs in a strip are
\ba\fl
V_2=\exp\bigg[(J'/k_{\mathrm{B}}T)\sum_{j=1}^{\bar p}\sigma^z_j\sigma^z_{j+1}\bigg]=
\exp\bigg[\ri(J'/k_{\mathrm{B}}T)\Big(\sum_{j=1}^{\bar p-1}
\Gamma_{2j}\Gamma_{2j+1}+U\Gamma_{1}\Gamma_{2\bar p}\Big)\bigg],
\label{V2}\ea
in which 
\be
U=\sigma_{1}^x\sigma_{2}^x\cdots\sigma_{\bar p}^x
=\ri^{\bar p}\Gamma_{1}\Gamma_{2}\cdots\Gamma_{2\bar p-1}\Gamma_{2\bar p}.
\ee
In the following we shall take the limit $\bar p\to\infty$ and only have operators from the even
sector of the Clifford algebra, that is the $\Gamma_j$'s only appear in pairs. Therefore, we
can ignore the $U$ replacing it by 1. The transfer matrices associated with vertical
interactions between the nearest-neighbor spin pairs are
\ba
V_1&&=\bigg[2\sinh(2J/k_{\mathrm{B}}T)]^{\bar p/2}
\exp\bigg[(J^*/k_{\mathrm{B}}T)\sum_{j=1}^{\bar p}\sigma^x_j\bigg]
\nonumber\\
&&=[2\sinh(2J/k_{\mathrm{B}}T)]^{\bar p/2}
\exp\bigg[\ri(J^*/k_{\mathrm{B}}T)\sum_{j=1}^{\bar p}\Gamma_{2j-1}\Gamma_{2j}\bigg],
\label{V1}\\
V_3&&=\bigg[2\sinh(2\hJ/k_{\mathrm{B}}T)]^{\bar p/2}
\exp\bigg[(\hJ^*/k_{\mathrm{B}}T)\sum_{j=1}^{\bar p}\sigma^x_j\bigg]
\nonumber\\
&&=[2\sinh(2\hJ/k_{\mathrm{B}}T)]^{\bar p/2}
\exp\bigg[\ri(\hJ^*/k_{\mathrm{B}}T)\sum_{j=1}^{\bar p}\Gamma_{2j-1}\Gamma_{2j}\bigg],
\label{V3}\ea
where $J^*$ and $\hJ^*$ are the dual variables of the Kramers--Wannier duality transform,
satisfying the relations
\be\fl
\sinh(2J/k_{\mathrm{B}}T)\sinh(2J^*/k_{\mathrm{B}}T)=1,\quad
\sinh(2\hJ/k_{\mathrm{B}}T)\sinh(2\hJ^*/k_{\mathrm{B}}T)=1.
\label{dual}\ee
As shown in \cite{HJPihd}, we can replace each string of length $n$ and nearest-neighbor
couplings $J$, by a single bond with coupling $\hJ$ obeying
\be
\tanh(\hJ/k_{\mathrm{B}}T)=\tanh^n(J/k_{\mathrm{B}}T)=z^n,
\quad z\equiv\tanh(J/k_{\mathrm{B}}T).
\label{thJ}\ee
The spin-spin correlations in the central rows of the strips%
\footnote{The central row of such a  strip is denoted as row zero.}
for $m$ even are then given by
\ba\fl
\langle \sigma_{0,1}\sigma_{0,\ell+1}\rangle=\Tr[\sigma_1\sigma_{\ell+1}\bT^p]/\Tr[\bT^p]
=\Tr\Bigg[\prod_{j=1}^{2\ell}(\ri\Gamma_{2j}\Gamma_{2j+1})\bT^p\Bigg]/\Tr[\bT^p],
\label{corr1}\ea
where
\be
\bT=V_2^{1/2}(V_1V_2)^{m/2}V_3(V_2V_1)^{m/2}V_2^{1/2},
\label{tm}\ee
with the $V_i$ defined in (\ref{V2}), (\ref{V1}) and (\ref{V3}). 
If the length $n$ of the strings is even, we can also calculate the row spin-pair correlations
for spins at the centers of a row of strings as
\ba\fl
\langle \sigma_{\bn,1}\sigma_{\bn,\ell+1}\rangle=\Tr[\sigma_1\sigma_{\ell+1}\bT'^p]/\Tr[\bT'^p]
=\Tr\Bigg[\prod_{j=1}^{2\ell}(\ri\Gamma_{2j}\Gamma_{2j+1})\bT'^p\Bigg]/\Tr[\bT'^p],
\label{corr1p}\ea
where
\be
\bT'=V_3^{1/2}(V_2V_1)^{ m}V_2V_3^{1/2},\quad \bn=(n+m)/2.
\label{htm}\ee

Using Wick's theorem, we may write (\ref{corr1}) and  (\ref{corr1p}) as
\ba
\langle \sigma_{0,1}\sigma_{0,\ell+1}\rangle=\underset{2\le j<k\le 2\ell+1}{\mathrm{Pf}} G(j,k),
\label{corr2}\\
\langle \sigma_{\bn,1}\sigma_{\bn,\ell+1}\rangle=\underset{2\le j<k\le 2\ell+1}{\mathrm{Pf}}G'(j,k),
\label{corr2p}\ea
where
\ba
G(j,k)=\Tr[(\ri\Gamma_{j}\Gamma_{k})\bT^p]/\Tr[\bT^p],
\label{Gp}\\
G(j,k)'=\Tr[(\ri\Gamma_{j}\Gamma_{k})\bT'^p]/\Tr[\bT'^p].\label{Gpp}\ea
We shall now concentrate ourselves on the analysis of (\ref{corr2}). The same steps
apply to (\ref{corr2p}) replacing $\bT$ by $\bT'$.

The complex conjugate of $G(j,k)$ is
\ba
\overline{G(j,k)}=\Tr[(\ri\Gamma_{j}\Gamma_{k})^\dagger(\bT^p)^\dagger]/\Tr[(\bT^p)^\dagger].
\ea
Using
\be
(\ri\Gamma_{j}\Gamma_{k})^\dagger=
-\ri\Gamma_{k}\Gamma_{j}=
\ri\Gamma_{j}\Gamma_{k}\quad \hbox{for $j\ne k$};
\quad (\bT^p)^\dagger=\bT^p,
\ee
we find for $j\ne k$
\be
\overline{G(j,k)}=G(j,k),\quad \overline{G(j,k)'}=G(j,k)'.
\label{gammap1}\ee
From (\ref{gamma}), we find for $j+k$ even, that $\Gamma_{j}\Gamma_{k}$ are matrices with real elements,
so that (\ref{gammap1}) implies $G(j,k)=0$, for $j\ne k$ and $j+k$ even. Consequently, in the limit
$\bar p\to\infty$, (\ref{corr2}) becomes the Toeplitz determinant
\ba
\langle \sigma_{0,1}\sigma_{0,\ell+1}\rangle=\left|\begin{array}{ccccc}
a_0&a_1&a_2&\cdots&a_{\ell-1}\\
a_{-1}&a_0&a_1&\cdots&a_{\ell-2}\\
a_{-2}&a_{-1}&a_0&\cdots&a_{\ell-3}\\
\vdots&\vdots&\vdots&\ddots&\vdots\\
a_{1-\ell}&a_{2-\ell}&a_{3-\ell}&\cdots&a_{0}
\end{array}\right|
=\det_{1\le j,k\le\ell} a_{k-j},
\label{Toeplitz}\ea
where 
\be
a_{k-j}=G(2j,2k+1).
\ee
A similar result holds for (\ref{corr2p}).

\section{Mathematical details}

Kaufman \cite{Kaufman} noted that the row-to-row transfer matrices $V_i$ are
$2^{\bar p}$-dimensional spinor representations of the group of (hyperbolic)
rotations in $2\bar p$ dimensions, as is expressed in full generality by the relations
\ba
V_i\equiv\exp\bigg(\half\ri\sum_{j=1}^{2\bar p}\sum_{k=1}^{2\bar p}
(\check\mathsf{C}_i)_{j,k}\Gamma_{j}\Gamma_{k}\bigg)\longleftrightarrow
\check\mathsf{V}_i\equiv\exp(2\ri\check\mathsf{C}_i),
\nonumber\\
V_i^{-1}\Gamma_{j}V_i=\sum_{k=1}^{2\bp}(\check\mathsf{V}_i)_{j,k}\Gamma_{k},\quad 1\le j\le 2\bp,
\label{cVi}\ea
where matrix $\check\mathsf{C}$ is antisymmetric.%
\footnote{A simple proof follows  working out the $t$-derivative after replacing
$\check\mathsf{C}$ by $t\check\mathsf{C}$ .}
At this point we could skip the next few steps citing \cite{Kaufman,KHuang,BaxIS}
and let the reader work out some of the minor differences for our case. Rather,
let us add some further details in order to be self-contained.

First, from (\ref{cVi}) we find  $\check\mathsf{V}_1$ and $\check\mathsf{V}_3$ to
be the block-diagonal matrices
\ba
\check\mathsf{V}_1=\left[\begin{array}{ccccc}
\bJ&0&0&\cdots&0\\
0&\bJ&0&\cdots&0\\
\vdots&\ddots&\ddots&\ddots&\vdots\\
0&\cdots&0&\bJ&0\\
0&\cdots&&0&\bJ
\end{array}\right]_{\bp\times\bp}\hspace{-1em},
\quad
\check\mathsf{V}_3=\left[\begin{array}{ccccc}
{\hat\bJ}&0&0&\cdots&0\\
0&{\hat\bJ}&0&\cdots&0\\
\vdots&\ddots&\ddots&\ddots&\vdots\\
0&\cdots&0&{\hat\bJ}&0\\
0&\cdots&&0&{\hat\bJ}
\end{array}\right]_{\bp\times\bp}\hspace{-1em},
\label{hV13}\ea
with
\ba
\bJ=\left[\begin{array}{cc}
\cosh(2J^*/k_{\mathrm{B}}T)&\ri\sinh(2J^*/k_{\mathrm{B}}T)\\
-\ri\sinh(2J^*/k_{\mathrm{B}}T)&\cosh(2J^*/k_{\mathrm{B}}T)\end{array}\right],
\label{bJ}\\
\hat\bJ=\left[\begin{array}{cc}
\cosh(2\hat J^*/k_{\mathrm{B}}T)&\ri\sinh(2\hat J^*/k_{\mathrm{B}}T)\\
-\ri\sinh(2\hat J^*/k_{\mathrm{B}}T)&\cosh(2\hat J^*/k_{\mathrm{B}}T)\end{array}\right].
\label{hbJ}\ea
In the limit $\bp\to\infty$, we choose $\check\mathsf{V}_2$ to be the block-cyclic
block-tridiagonal matrix\footnote{%
For finite  $\bp$ we also need the block-anti-cyclic version giving
the same result in the limit $\bp\to\infty$. More precisely, we  insert
$\mathbf{1}=\half(\mathbf{1}+U)+\half(\mathbf{1}-U)$ within each trace
in (\ref{corr1}) and (\ref{corr1p}) and apply (35) of  \cite{Kaufman} to (\ref{V2}).
In the limit $\bp\to\infty$, the two terms with $U$ cancel and the other two become equal.}
\ba
\check\mathsf{V}_2=\left[\begin{array}{ccccc}
\bC_2&\bB_2&0&\cdots&\bB_2^\dagger\\
\bB_2^\dagger&\bC_2&\bB_2&\cdots&0\\
\vdots&\ddots&\ddots&\ddots&\vdots\\
0&\cdots&\bB_2^\dagger&\bC_2&\bB_2\\
\bB_2&0&\cdots&\bB_2^\dagger&\bC_2
\end{array}\right],
\label{hV2}\ea
in which
\be\fl
\bC_2=\left[\begin{array}{cc}
\cosh(2J'/k_{\mathrm{B}}T)&0\\
0&\cosh(2J'/k_{\mathrm{B}}T)\end{array}\right],\quad
\bB_2=\left[\begin{array}{cc}
0&0\\
\ri\sinh(2J'/k_{\mathrm{B}}T)&0\end{array}\right].
\label{hV2a}\ee
The matrix $\check\mathsf{V}_2$ in (\ref{hV2}) may block diagonalized by
discrete Fourier transform that leaves the block-diagonal $\check\mathsf{V}_1$
and $\check\mathsf{V}_3$ invariant. More specifically,
let the elements of the $2\bp\times2\bp$ matrix $\bcP$ and its inverse be
\be\fl
\bcP_{rs}=\re^{\ri r\theta_s}\mathbf {1}_2/\sqrt{\bp};\quad \bcP^{-1}_{st}=
\re^{-\ri t\theta_s}\mathbf {1}_2/\sqrt{\bp};
\quad \theta_s=2\pi s/\bp,\quad r,s,t=0,1\cdots\bp-1,
\label{CP}\ee
with $\mathbf {1}_2$ the $2\times2$ unit matrix. Then
\be\bcP^{-1} \check\mathsf{V}_2\bcP=\left[\begin{array}{ccccc}
\bA_2(\theta_0)&0&0&\cdots&0\\
0&\bA_2(\theta_1)&0&\cdots&0\\
\vdots&\ddots&\ddots&\ddots&\vdots\\
0&\cdots&0&\bA_2(\theta_{\bp-2})&0\\
0&\cdots&&0&\bA_2(\theta_{\bp-1})
\end{array}\right],
\label{hV2d}\ee
where
\ba
\bA_2(\theta)=\bC_2+\bB_2\re^{\ri\theta}+\bB^{\dagger}_2\re^{-\ri\theta}\nonumber\\
=\left[\begin{array}{cc}
\cosh(2J'/k_{\mathrm{B}}T)&-\ri\sinh(2J'/k_{\mathrm{B}}T)\re^{-\ri\theta}\\
\ri\sinh(2J'/k_{\mathrm{B}}T)\re^{i\theta}&\cosh(2J'/k_{\mathrm{B}}T)
\end{array}\right].
\label{bA2}\ea
Now from (\ref{tm}) and using (\ref{cVi}), (\ref{hV13}) and (\ref{hV2d}) we find
\be \bT^{-1}\Gamma_{j}\bT=\sum_{\ell}(\check \bT)_{j,\ell}\Gamma_{\ell},
\label{TGT}\ee
with
\be
\bcP^{-1}{\check\bT}\,\bcP=\left[\begin{array}{ccccc}
\bA(\theta_0)&0&0&\cdots&0\\
0&\bA(\theta_1)&0&\cdots&0\\
\vdots&\ddots&\ddots&\ddots&\vdots\\
0&\cdots&0&\bA(\theta_{\bp-2})&0\\
0&\cdots&&0&\bA(\theta_{\bp-1})
\end{array}\right],
\label{phV2}\ee
in which
\be
\bA(\theta)=\bA_2(\theta)^{\halfs}[\bJ\bA_2(\theta)]^{\halfs m}{\hat\bJ}
[\bA_2(\theta)\bJ]^{\halfs m}\bA_2(\theta)^{\halfs},
\label{bA}\ee
as seen from  (\ref{tm}), and when (\ref{phV2}) is used. 

Using (\ref{GG}), we may write (\ref{Gp}) as%
\footnote{The next few steps are similar to those in (2.17)--(2.20) in \cite{McCoyPerk}
and those following (4.19) in \cite{PerkCapel}.}
\be
G(j,k)+G(k,j)=2\ri\delta_{j,k}.
\label{GpG}\ee
Rewriting (\ref{Gp}) and applying (\ref{TGT}), we find
\be
G(k,j)=\Tr[(\ri\Gamma_{k}\bT^p(\bT^{-p}\Gamma_{j})\bT^p)]/\Tr[\bT^p]
=\sum_\ell {\check\bT}^p_{j,\ell}G(\ell,k),
\ee
so that (\ref{GpG}) becomes
\be
G(j,k)+\sum_\ell {\check\bT}^p_{j,\ell}G(\ell,k)=2\ri\delta_{j,k}.
\label{GpG2}\ee
Consequently, if we let $\bG$ denote the $2\bp\times2\bp$ matrix whose
elements are $G(k,j)$, then the above equation can be rewritten as
\be
\bG+{\check\bT}^p\bG=2\ri{\bf 1}_{2\bp},
\ee
or equivalently
\be
\bcP^{-1}\bG\bcP({\bf 1}_{2\bp}+\bcP^{-1}{\check\bT}^p\bcP)=2\ri{\bf 1}_{2\bp}.
\label{pGT}\ee
Due to the cyclic boundary condition, we have
$G(j,k)=G(j+2\ell,k+2\ell)=G(j+2\bp,k)=G(j,k+2\bp)$.
Therefore, the matrix $\bG$ is also $2\times2$ block-cyclic, and can
be diagonalized by the matrices in (\ref{CP}) as
\be
(\bcP^{-1}\bG\bcP)_{\ell,m}=\delta_{\ell,m}\tilde{\bG}(\theta_{\ell}).
\label{dbG}\ee
Since, from the text following (\ref{gammap1}), we have $G(j,j)=\ri$ and $G(j+2k,j)=0$,
we find that the Fourier transform $\tilde{\bG}(\theta)$ is a $2\times2$ matrix given by
\be \tilde{\bG}(\theta)
=\left[\begin{array}{cc}
\ri&\tilde{\bG}(\theta)_{12}\\
\tilde{\bG}(\theta)_{21}&\ri\end{array}\right],
\label{FourG}\ee
in which 
\be\tilde{\bG}(\theta)_{12}=\sum_{k=0}^{\bp-1}\re^{\ri\theta k}{\bG}(1,2+2k),\quad
\tilde{\bG}(\theta)_{21}=\sum_{k=0}^{\bp-1}\re^{\ri\theta k}{\bG}(2,1+2k).
\label{FourG12}\ee

Because of the block diagonal forms given in (\ref{phV2}) and (\ref{dbG}),
we may rewrite (\ref{pGT}) as
\be
\tilde{\bG}(\theta)
=2\ri{\bf 1}_2\bigg/[{\bf 1}_2+\bA(\theta)^p].
\label{GrA}\ee
We shall now diagonalize the matrix $\bA(\theta)$ given in (\ref{bA}).  Using
standard notations \cite{BaxIS,MWbk} $z=\tanh(J/k_{\mathrm{B}}T)$,
$z'=\tanh(J'/k_{\mathrm{B}}T)$, $z^{\ast}=\tanh(J^{\ast}/k_{\mathrm{B}}T)$, so that
\ba
\cosh\frac{2J'}{k_{\mathrm{B}}T}=\frac{1+z'{}^2}{1-z'{}^2},\quad
\sinh\frac{2J'}{k_{\mathrm{B}}T}=\frac{2z'}{1-z'{}^2},\nonumber\\
\cosh\frac{2J^*}{k_{\mathrm{B}}T}=\frac{1+{z^*}^2}{1-{z^*}^2},\quad
\sinh\frac{2J^*}{k_{\mathrm{B}}T}=\frac{2z^*}{1-{z^*}^2},\quad z^*=\frac{1-z}{1+z},
\label{JpJstar}\ea
we find 
\ba\fl
\bJ\bA_2(\theta)=\bY\left[\begin{array}{cc}
\alpha(\theta)&0\\
0&\alpha(\theta)^{-1}\end{array}\right]\bY^{-1},
\quad
\bA_2(\theta)\bJ=\bX\left[\begin{array}{cc}
\alpha(\theta)&0\\
0&\alpha(\theta)^{-1}\end{array}\right]\bX^{-1}
\label{A2J}\ea
with real positive eigenvalues
\ba
\alpha(\theta)^{\pm 1}=g(\theta)\pm\sqrt{g^2(\theta)-1},\nonumber\\
 g(\theta)= [(1+{z'}^2)(1+{z^*}^2)-4z'z^*\cos\theta]\Big/[(1-{z'}^2)(1-{z^*}^2)].
\label{G} \ea
It can be easily verified that
 \ba
 {g(\theta)^2-1}&=&\frac{4(1-z'z^*\re^{\ri\theta})(1-z'z^*\re^{-\ri\theta})(z'-z^*\re^{\ri\theta})
 (z'-z^*\re^{-\ri\theta})}{(1-{z'}^2)^2(1-{z^*}^2)^2}\label{G2}\\
 &=&\Bigg[\frac{2(1-z'z^*\re^{-\ri\theta})(z'-z^*\re^{\ri\theta})}{(1-{z'}^2)(1-{z^*}^2)}\Bigg]^2\Omega,
 \label{GOmega}\ea
\be
\Omega=\frac{(1-z'z^*\re^{\ri\theta})(z'-z^*\re^{-\ri\theta})}
{(1-z'z^*\re^{-\ri\theta})(z'-z^*\re^{\ri\theta})},\qquad \Omega^{-1}=\overline\Omega,
\label{Omega}\ee
and
\ba
\bY=\left[\begin{array}{cc}
1-z'\sqrt{\Omega}&\ri\re^{-\ri\theta}(1+z'\sqrt{\Omega})\\
-\ri\re^{\ri\theta}(z'-\sqrt{\Omega})&(z'+\sqrt{\Omega})\end{array}\right],
\nonumber\\
\bX=\left[\begin{array}{cc}
1+z'\sqrt{\Omega}&\ri\re^{-\ri\theta}(1-z'\sqrt{\Omega})\\
\ri\re^{\ri\theta}(z'+\sqrt{\Omega})&-(z'-\sqrt{\Omega})\end{array}\right], \quad 
\bA_2(\theta)\bY=\bX,
\label{bYX}\ea
provided we choose the square roots $\sqrt{g^2(\theta)-1}$ and $\Omega^{1/2}$
to be positive for $\theta=\pm\pi$.

Comparing the duality relation (\ref{dual}) with (\ref{thJ}) we see
\be
\tanh(\hJ/k_{\mathrm{B}}T)=z^n=\re^{-2\hJ^*/k_{\mathrm{B}}T},
\hJ^*=n J^*,
\label{dual2}\ee
consistent with $V_3={V_1}^n$,
so that elements of matrix $\hat\bJ$ given by (\ref{hbJ}) can be written as
\be
\cosh(2\hJ^*/k_{\mathrm{B}}T)=\half(z^{-n}\!+\!z^n),\quad
\sinh(2\hJ^*/k_{\mathrm{B}}T)=\half (z^{-n}\!-\!z^n).
\ee
Now using (\ref{hbJ}) and (\ref{bYX}), we can evaluate 
\ba
\bY^{-1}{\hat\bJ}\bX=\left[\begin{array}{cc}
p_{00}&\ri\re^{-\ri\theta}p_{01}\\
\ri\re^{\ri\theta}p_{01}&p_{11}\end{array}\right],\quad p_{00}p_{11}+p^2_{01}=1,
\label{YhJX}\ea
with
\ba\fl
p_{00}=\frac{(z^{-n}\!+\!z^n)(z'\!+\!\sqrt{\Omega})(1\!+\!z'\sqrt{\Omega})
-\half(z^{-n}\!-\!z^n)[\re^{\ri\theta}(z'\!+\!\sqrt{\Omega})^2
+\re^{-\ri\theta}(1\!+\!z'\sqrt{\Omega})^2]}{2\sqrt{\Omega}(1-z'{}^2)},
\nonumber\\
\fl
p_{11}=\frac{-(z^{-n}\!+\!z^n)(z'\!-\!\sqrt{\Omega})(1\!-\!z'\sqrt{\Omega})
+\half(z^{-n}\!-\!z^n)[\re^{\ri\theta}(z'\!-\!\sqrt{\Omega})^2
+\re^{-\ri\theta}(1\!-\!z'\sqrt{\Omega})^2]}{2\sqrt{\Omega}(1-z'{}^2)},
\nonumber\\
\fl
p_{01}=\frac{(z^{-n}\!+\!z^n)z'(1-\Omega)-\half(z^{-n}\!-\!z^n)
[\re^{\ri\theta}(z'{}^2-{\Omega})+\re^{-\ri\theta}(1-z'{}^2{\Omega})]}{2\sqrt{\Omega}(1-z'{}^2)}.
\label{p}\ea
We note that $\overline{p_{00}}=p_{00}$, $\overline{p_{11}}=p_{11}$ and
$\overline{p_{01}}=-p_{01}$, as complex conjugation replaces $\sqrt{\Omega}$ by its inverse;
so $p_{00}$ and  $p_{11}$ are real and  $p_{01}$ is imaginary.

We rewrite (\ref{bA}) as
\be\fl
\bA(\theta)=\left[\bA_2(\theta)^{1/2}\bY\right]\!\left[\bY^{-1}\bJ\bA_2(\theta)\bY\right]^{m/2}\!
\big[\bY^{-1}{\hat\bJ}\bX\big]\!
\left[\bX^{-1}\bA_2(\theta)\bJ\bX\right]^{m/2}\!\left[\bX^{-1}\bA_2(\theta)^{1/2}\right],
\label{bAnew}\ee
where the middle three factors have been given in (\ref{A2J}) and (\ref{YhJX}).
Now $\bA_2(\theta)^{1/2}$ is given by (\ref{bA2}) with $2J'$ replaced by $J'$
and $\bX$ and $\bY$ are given in (\ref{A2J}) with $z'=\tanh(J'/k_{\mathrm{B}}T)$.
Therefore, we can simplify the remaining two factors in (\ref{bAnew}) as
\ba
\bA_2(\theta)^{1/2}\bY=\frac{1}{\cosh(J'/k_{\mathrm{B}}T)}
\left[\begin{array}{cc}
1&\ri\re^{-\ri\theta}\\
\ri\re^{\ri\theta}\sqrt{\Omega}&\sqrt{\Omega}
\end{array}\right],
\nonumber\\
\bX^{-1}\bA_2(\theta)^{1/2}=\frac{\cosh(J'/k_{\mathrm{B}}T)}{2}
\left[\begin{array}{cc}
1&-\ri\re^{-\ri\theta}/\sqrt{\Omega}\\
-\ri\re^{\ri\theta}&1/\sqrt{\Omega}
\end{array}\right].
\ea
Multiplying the five matrix factors in (\ref{bAnew}), we obtain
\ba
\bA(\theta)=\left[\begin{array}{cc}
W&-\ri\re^{-\ri\theta}Z/\sqrt{\Omega}\\
\ri\re^{\ri\theta}\overline{Z}\,\sqrt{\Omega}&W\end{array}\right].
\label{bAf}\ea
in which 
\be
W=\half(p_{00}\alpha^m+p_{11}\alpha^{-m}),\quad
Z=\half(p_{00}\alpha^m-p_{11}\alpha^{-m})-p_{01}.
\label{WZ}\ee
As $\bA(\theta)$ derives from the complex rotations (\ref{bJ}), (\ref{hbJ}),
(\ref{hV2d}) and (\ref{bA2}), its determinant should be 1. Indeed, also
from (\ref{bAf}) and (\ref{YhJX}) it follows that
\be |\bA(\theta)|=W^2-Z\overline{Z}=1,\ee
so that the eigenvalues of $\bA(\theta)$ are $\Lambda^{\pm1}$, satisfying
\be
|\bA(\theta)-\Lambda|=0=(W-\Lambda)^2-Z\overline{Z}=\Lambda^2-2W\Lambda+1,
\ee
or
\be
\Lambda^{\pm1}=W\pm\sqrt{W^2-1}=W\pm\sqrt{Z\overline{Z}}.
\ee
Matrix $\bA(\theta)$ is hermitian and its normalized eigenvectors are
\ba
v_{\pm1}=2^{-1/2}\Big[\!\begin{array}{c}\pm x\\1\end{array}\!\Big],\quad
v_i^{\dagger}v_j^{\vphantom{\dagger}}=\delta_{ij},\;(i,j=\pm1),
\nonumber\\
x=-\frac{\ri\re^{-\ri\theta}}{\sqrt{\Omega}}\sqrt{\frac Z{\overline{Z}}},\quad
\overline{x}=x^{-1},\quad
\bA(\theta)=\sum_{i=\pm1}\Lambda^{i}v_i^{\vphantom{\dagger}}v_i^{\dagger}.
\ea
Therefore,
\ba
\bA(\theta)^p&=&2^{-1}\left[\begin{array}{cc}
x&-x\\1&1\end{array}\right]
\left[\begin{array}{cc}
\Lambda^p&0\\
0&\Lambda^{-p}\end{array}\right]
\left[\begin{array}{cc}
\overline{ x}&1\\-\overline{ x}&1\end{array}\right]\nonumber\\
&=&\left[\begin{array}{cc}
\half(\Lambda^p+\Lambda^{-p})
&\half x(\Lambda^p-\Lambda^{-p})\\
\half\overline{x}(\Lambda^p-\Lambda^{-p})&\half(\Lambda^p+\Lambda^{-p})
\end{array}\right].
\label{eigenbA}\ea

Consequently, we have
\be
|1+\bA(\theta)^p|=2+\Lambda^p+\Lambda^{-p}
\ee
and
\ba\fl
[1+\bA(\theta)^p]^{-1}=\frac 1{2+\Lambda^p+\Lambda^{-p}}
\left[\begin{array}{cc}
1+\half(\Lambda^p+\Lambda^{-p})
&-\half x(\Lambda^p-\Lambda^{-p})\\
-\half\overline{ x}(\Lambda^p-\Lambda^{-p})&1+\half(\Lambda^p+\Lambda^{-p})
\end{array}\right].
\label{bGmatrix0}\ea
In the limit $p\to\infty$, we find from (\ref{GrA}), that
\be
\tilde{\bG}(\theta)=\left[\begin{array}{cc}
\ri&-\ri x\\
-\ri\overline{ x}&\ri\end{array}\right]
\label{bGmatrix}\ee
so that
\be\fl
\tilde{\bG}(\theta)_{12}=-\ri x=-\frac{\re^{-\ri\theta}}{\sqrt{\Omega}}\sqrt{
\frac Z{\overline{Z}}},
\quad
\tilde{\bG}(\theta)_{21}=-\ri{\overline x}=
{\re^{\ri\theta}}{\sqrt{\Omega}}\sqrt{\frac {\overline{Z}}Z}.
\label{G12}\ee
Therefore, we can then use (\ref{FourG12}) to write
\ba
{\bG}(2,1+2\ell+2)={\bG}(2k,1+2\ell+2k)=a_{\ell}
\nonumber\\
=\frac1\bp\sum_{j=0}^{\bp-1}\re^{-\ri\theta_j (\ell+1)}\tilde{\bG}(\theta)_{21}\to
\frac1{2\pi}\int_{0}^{2\pi}\rd \theta\, \re^{-\ri \ell\theta}\,{\sqrt{\Omega}}\sqrt{
{\overline{Z}}/Z}.
\ea
in the limit $\bp\to\infty$.
Now we are going to analyze $Z$ as given by (\ref{WZ}). Using  (\ref{p}),
and setting $m=2j$ we may write
\ba\fl
[8\sqrt{\Omega}(1\!-\!z'{}^2)]Z =
z^{-n}\re^{-\ri\theta}\{(\alpha^m+\alpha^{-m}-2)[z^{2n}(1+z'\re^{\ri\theta})^2-(1-z'\re^{\ri\theta})^2]\nonumber\\
+\Omega(\alpha^m+\alpha^{-m}+2)[z^{2n}(z'+\re^{\ri\theta})^2-(z'-\re^{\ri\theta})^2]
\nonumber\\
+2\sqrt{\Omega}(\alpha^m-\alpha^{-m})
[z^{2n}(z'+\re^{\ri\theta})(1+z'\re^{\ri\theta})-(z'-\re^{\ri\theta})(1-z'\re^{\ri\theta})]\}
\nonumber\\
\fl=z^{-n}\re^{-\ri\theta}\{z^{2n}[(1+z'\re^{\ri\theta})(\alpha^{\halfs m}\!-\alpha^{-\halfs m})
+\sqrt{\Omega}(z'+\re^{\ri\theta})(\alpha^{\halfs m}\!+\alpha^{-\halfs m})]^2
\nonumber\\
-[(1-z'\re^{\ri\theta})(\alpha^{\halfs m}\!-\alpha^{-\halfs m})+
\sqrt{\Omega}(z'-\re^{\ri\theta})(\alpha^{\halfs m}\!+\alpha^{-\halfs m})]^2\}.
\ea
This shows that $Z$ can be factorized. Letting $m=2j$ and defining
\ba
A(\theta)=&(\alpha^j+\alpha^{-j})[z'(z^n-1)\re^{-\ri\theta}+(z^n+1)]
\nonumber\\
&+\Omega^{-\halfs}(\alpha^j-\alpha^{-j})[(z^n-1)\re^{-\ri\theta}+z'(z^n+1)],
\nonumber\\
B(\theta)=&(\alpha^j+\alpha^{-j})[(z^n-1)\re^{\ri\theta}+z'(z^n+1)]
\nonumber\\
&+\Omega^{-\halfs}(\alpha^j-\alpha^{-j})[z'(z^n-1)\re^{\ri\theta}+(z^n+1)],
\label{ABp}\ea
we find
\be 
Z=\sqrt{\Omega}A(\theta)B(\theta)/[8z^n(1\!-\!z'{}^2)],\quad 
{\overline Z}={\overline A(\theta)}{\overline B(\theta)})/[8z^n(1\!-\!z'{}^2)\sqrt{\Omega}].
\ee
Consequently, we have
\ba 
a_n=\frac 1{2\pi}\int_{-\pi}^{\pi}\rd \theta\, \re^{-\ri n\theta}\,\Phi(\theta),
\quad \Phi(\theta)={\sqrt{\Omega}}\sqrt{
\frac{\overline{Z}}Z}
=\sqrt\frac{\overline {A(\theta)}\,\overline {B(\theta)}}{A(\theta)B(\theta)}.
\label{Phi}\ea 

As for the spin-pair correlations (\ref{corr2p}) for spins at the centers of a row of strings,
we need to replace $\bT$ by $\bT'$. Here we shall briefly outline the differences.
Similar to (\ref{phV2}), we find
\be 
{\bT'}^{-1}\Gamma_{j}\bT'=\sum_{k}(\check {\bT'})_{j,k}\Gamma_{k}\ee
with 
\be
\bcP^{-1}{ \check\bT'}\,\bcP=\left[\begin{array}{ccccc}
\bA'(\theta_0)&0&0&\cdots&0\\
0&\bA'(\theta_1)&0&\cdots&0\\
\vdots&\ddots&\ddots&\ddots&\vdots\\
0&\cdots&0&\bA'(\theta_{\bp-2})&0\\
0&\cdots&&0&\bA'(\theta_{\bp-1})
\end{array}\right].
\ee
As seen from (\ref{htm}), we have
\be
\bA'(\theta)={\hat\bJ}^{\halfs}\bA_2(\theta)[\bJ\bA_2(\theta)]^{m}{\hat\bJ}^{\halfs}.
\label{bAp}\ee
Relation (\ref{pGT}) also holds with matrix ${\bG}$ replaced by ${\bG'}$ of (\ref{Gp})
and $ {\check\bT}$ replaced by $ {\check\bT'}$. Therefore, due to the cyclic boundary
condition of ${\bG'}$, we find, similar to  (\ref{GrA}), that
\be
\tilde{\bG'}(\theta)
=2\ri{\bf 1}_2\bigg/[{\bf 1}_2+\bA'(\theta)^p].
\label{GrAp}\ee
Using (\ref{A2J}) and  (\ref{bYX}), we rewrite (\ref{bAp}) as
\ba
\bA'(\theta)={\hat\bJ}^{\halfs}\bX\left[\begin{array}{cc}
\alpha^m&0\\
0&\alpha^{-m}\end{array}\right]\bY^{-1}{\hat\bJ}^{\halfs}
\nonumber\\
=\left[\begin{array}{cc}
W&{-\ri\e^{-\ri\theta}}B(\theta)\overline{A(\theta)}/[8z^n(1-z'{}^2)]\\
{\ri\re^{\ri\theta}}\overline{B(\theta)}A(\theta)/[8z^n(1-z'{}^2)]&W
\end{array}\right],
\ea
where $W$ is defined in (\ref{WZ}), and $A(\theta)$ and $B(\theta)$ are
defined in (\ref{ABp}). As $|\bA'(\theta)|=1$,
it is easily seen that $\bA'(\theta)$ has same eigenvalues as $\bA(\theta)$,
but not the same eigenvectors. Using the
same steps as in (\ref{eigenbA}) to (\ref{bGmatrix}), we find that
\be
\tilde{\bG'}(\theta)_{12}=\re^{-\ri\theta}\sqrt{
\frac{\overline{A(\theta)}B(\theta)}{A(\theta)\overline{B(\theta)}}},
\quad
\tilde{\bG'}(\theta)_{21}=-\re^{\ri\theta}\sqrt{
\frac{A(\theta)\overline{B(\theta)}}{\overline{A(\theta)}B(\theta)}},
\label{G12p}\ee
so that
\ba
a'_n=\frac 1{2\pi}\int_{-\pi}^{\pi}\rd \theta\, \re^{-\ri n\theta}\,\Phi'(\theta),
\quad \Phi'(\theta)=
\sqrt\frac{{A(\theta)}\,\overline {B(\theta)}}{\overline{A(\theta)}\,B(\theta)}.
\label{Phip}\ea
We shall now examine some limiting cases for the generating function.

\subsection{Case 1: $m=2j\to\infty$:}

In the limit $m=2j\to\infty$, we may drop $\alpha^{-j}$ in (\ref{ABp}), and find
\be
\overline{A(\theta)}=\Omega^{\halfs}B(\theta),\quad
 \overline{B(\theta)}=\Omega^{\halfs}A(\theta),
 \ee
so that the generating function in (\ref{Phi}) becomes
 \be
 \Phi(\theta)=\sqrt\frac{\overline {A(\theta)}\,\overline {B(\theta)}}{A(\theta)B(\theta)}
 =\Omega^{\halfs},\label{PhiMW}\ee
which is identical to the generating function given in (1.3) and (1.4) on page 249
in McCoy and Wu's book \cite{MWbk} for the row correlation, as it should.

For the generating function in (\ref{Phip}) for spins at the center of a string row,
 \be
 \Phi'(\theta)=\sqrt\frac{{A(\theta)}\overline {B(\theta)}}{\overline{A(\theta)}B(\theta)}
 =\frac{A(\theta)}{B(\theta)}=
 \frac{(z'-z_n^*\re^{-\ri\theta})+\Omega^{1/2}(1-z'z_n^*\re^{-\ri\theta})}
 {(1-z'z_n^*\re^{\ri\theta})+\Omega^{1/2}(z'-z_n^*\re^{\ri\theta})},
 \label{Phipm}\ee
 where $z_n^*=(1-z^n)/(1+z^n)=\tanh(nJ^*/k_{\mathrm{B}}T)$, with $J^*$ the dual
 of $J$ at given $T$ defined in (\ref{dual}), see also (\ref{dual2}). We expect the
resulting pair correlation to have continuously varying critical exponents, as this
correlation can be equivalently calculated from a infinite uniform Ising model with
one whole horizontal row of vertical bonds $J$ replaced by Ising chains of length
$n$ with the same $J$. These chains are equivalent to single bonds $\hat J$
(\ref{thJ}). But that makes the model one of the dual pair of linear defect models of 
Bariev \cite{Bariev}.

\subsection{Case 2: $n=1$:}

For $n=1$, it is the regular Ising model, so that the generating function should be
same as (\ref{PhiMW}). Setting $n=1$ in (\ref{ABp}) and using (\ref{JpJstar}), we find
\ba\fl
A(\theta)=(z+1)[(\alpha^j+\alpha^{-j})(1-z'z^*\re^{-\ri\theta})+
\Omega^{-\halfs}(\alpha^j-\alpha^{-j})(z'-z^*\re^{-\ri\theta})]
\nonumber\\
=(z+1)(1-z'z^*\re^{-\ri\theta})[(\alpha^j+\alpha^{-j})+(\alpha^j-\alpha^{-j})\Delta^{\halfs}],
\label{An1}\ea
From (\ref{Omega}), we find
\be
\Delta^{\halfs}=\Omega^{-\halfs}\frac{(z'-z^*\re^{-\ri\theta})}{(1-z'z^*\re^{-\ri\theta})}
=\left[\frac{(z'-z^*\re^{\ri\theta})(z'-z^*\re^{-\ri\theta})}
{(1-z'z^*\re^{\ri\theta})(1-z'z^*\re^{-\ri\theta})}\right]^\halfs,
\ee
which is real, $\Delta=\overline\Delta$. Consequently, we have
\be
\overline{A(\theta)}/A(\theta)=(1-z'z^*\re^{\ri\theta})/(1-z'z^*\re^{-\ri\theta}),
\ee
Similarly we find that
\ba
B(\theta)=(z+1)(z'-z^*\re^{\ri\theta})[(\alpha^j+\alpha^{-j})+\Delta^{-\halfs}(\alpha^j-\alpha^{-j})],
\label{Bn1}\ea
resulting in
\be
\overline{B(\theta)}/B(\theta)=(z'-z^*\re^{-\ri\theta})/(z'-z^*\re^{\ri\theta}).
\ee
Thus the generating function for $n=1$ is the one of the regular Ising model,
 \be
 \Phi(\theta)= \sqrt\frac{\overline {A(\theta)}\,\overline {B(\theta)}}{A(\theta)B(\theta)}
 =\sqrt{\re^{-2\ri\theta}\frac{\overline{A(\theta)}^2}{A(\theta)^2}}=\Omega^{\halfs}.
 \ee
 For $n=1$, the strings are of length 1 and do not have center rows.

\subsection{Case 3: $n=\infty$:}

In the limit $n\to\infty$, we have $z^n\to0$, and (\ref{ABp}) becomes
 \ba
A(\theta)=(\alpha^j+\alpha^{-j})(1-z'\re^{-\ri\theta})+
\Omega^{-\halfs}(\alpha^j-\alpha^{-j})(z'-\re^{-\ri\theta}),\nonumber\\
B(\theta)=(\alpha^j+\alpha^{-j})(z'-\re^{\ri\theta})+
\Omega^{-\halfs}(\alpha^j-\alpha^{-j})(1-z'\re^{\ri\theta}).\label{ABz}\ea
Therefore,
 \be
 B(\theta)\to-\re^{\ri\theta}A(\theta),\qquad
\overline{B(\theta)}\to-\re^{-\ri\theta}\overline{A(\theta)},
\label{ABzn} \ee
 so that
 \be
 \Phi(\theta)=\sqrt{\re^{-2\ri\theta}\frac{\overline{A(\theta)}^2}{A(\theta)^2}}=-\re^{-\ri\theta}\frac{\overline{A(\theta)}}{A(\theta)}.
 \label{Phiz0}\ee
The choice of sign is to make $-\re^{-\ri\pi}=1$. Because all square roots disappear, 
also in the product $\Omega^{-\halfs}(\alpha^j-\alpha^{-j})$ as can be seen from
(\ref{G})--(\ref{Omega}), the correlation function determined by (\ref{Phiz0}) behaves
very differently from 2-d Ising, decaying exponentially as in the one-dimensional
Ising model. More precisely, the pair correlation is identical to the one in the middle row of an
infinite strip of width $m=2j$ with free boundaries.

Using (\ref{ABzn}), we find for $n=\infty$, that the generating function in (\ref{Phip}) for
the spin-pair correlation in a row of the string centers becomes
 \be
 \Phi'(\theta)=\sqrt\frac{A(\theta)\overline {B(\theta)}}{\overline {A(\theta)}B(\theta)}
 =-\re^{-\ri\theta},
 \label{Phiz0p}\ee
 so that $a'_l=-\delta_{l,-1}$ in (\ref{Toeplitz}) and the spins in that row are
 uncorrelated, as to be expected.
 
\section{Spontaneous magnetizations}

We shall now use Szeg\H o's theorem to calculate the spontaneous magnetizations.
From (\ref{ABp}) and (\ref{GOmega}), we can see that the $A(\theta)$ and $B(\theta)$
as functions of $\re^{i\theta}$ have $2j+1$ roots.
They can only be calculated numerically. From these calculations, we find that all
the roots are real; and $A(\theta)$ has $j+1$ roots
smaller than 1, and $j$ roots greater than 1 for all temperatures,
while $B(\theta)$ has $j+1$ roots smaller than 1 and $j$ roots greater
than 1 for $T>T_{\mathrm c}(1,m,n)$, but one of the roots, say $\gamma_{j+1}$,
becomes 1 at the critical temperature, and greater
than 1 for $T<T_{\mathrm c}(1,m,n)$. We rewrite (\ref{ABp}) as 
\ba
A(\theta)=\rho_a\prod_{\ell=1}^{j+1}(1-{\hat\gamma}_\ell\re^{-\ri\theta} )
               \prod_{\ell=j+2}^{2j+1}(1-{\hat\gamma}^{-1}_\ell\re^{\ri\theta} ),\nonumber\\
B(\theta)=\rho_b\prod_{\ell=1}^{j}(1-{\gamma}_\ell\re^{-\ri\theta} )
               \prod_{\ell=j+1}^{2j+1}(1-{\gamma}^{-1}_\ell\re^{\ri\theta} ),
\label{AB}\ea
where $\rho_a$ and $\rho_b$ are real constants. Having these roots we
can apply Szeg\H o's theorem:
\ba
{M}^2=\lim_{r\to\infty}\langle \sigma_{0,1}\sigma_{0,r+1}\rangle=
\exp\bigg(\sum_{n=1}^\infty n g_n g_{-n}\bigg),
\label{mag}\ea
where 
\ba
g_n=\frac 1{2\pi}\int_{0}^{2\pi}\rd \theta\, \re^{-\ri n\theta}\,\ln\Phi(\theta),
\label{mgn}\ea
provided $\re^{g_0}=1$. 

\subsection{Spontaneous magnetization at the center row of the strip}
For the spontaneous magnetization at the center of the strip, we substitute
(\ref{Phi}) into (\ref{mgn}) and use (\ref{AB}) to
find that $g_0=0$ and
\ba
g_n=- g_{-n}=\frac 1{2n}\left[\sum_{\ell=1}^{j+1}{\hat\gamma}^n_\ell-
\sum_{\ell=j+2}^{2j+1}{\hat\gamma}^{-n}_\ell
+\sum_{\ell=1}^{j}\gamma^n_\ell-\sum_{\ell=j+1}^{2j+1}\gamma^{-n}_\ell\right],
\label{mgn1}\ea
such that
\ba
\sum_{n=1}^\infty n g_n g_{-n}=-\sum_{n=1}^\infty \frac 1 {4n}
\left[\sum_{\ell=1}^{j+1}{\hat\gamma}^n_\ell-\sum_{\ell=j+2}^{2j+1}
{\hat\gamma}^{-n}_\ell+\sum_{\ell=1}^{j}\gamma^n_\ell-
\sum_{\ell=j+1}^{2j+1}\gamma^{-n}_\ell\right]^2
\nonumber\\
\hspace{60pt}=\ln[(1-\gamma^{-2}_{j+1})^{1/4}{\mathcal{F}}\mathcal{H}],
\label{ngg}\ea
in which
\ba\fl
\ln[(1-\gamma^{-2}_{j+1})^{1/4}\mathcal{F}]=
-\sum_{n=1}^\infty \frac 1 {4n}\Bigg[\bigg(\sum_{\ell=1}^{j+1}{\hat\gamma}^n_\ell\bigg)^2
+\bigg(\sum_{\ell=j+2}^{2j+1}{\hat\gamma}^{-n}_\ell\bigg)^2
+\bigg(\sum_{\ell=1}^{j}\gamma^n_\ell\bigg)^2
+\bigg(\sum_{\ell=j+1}^{2j+1}\gamma^{-n}_\ell\bigg)^2
\nonumber\\
\hspace{75pt}
-2\sum_{\ell=1}^{j+1}
\sum_{k=j+2}^{2j+1}({\hat\gamma}_\ell/{\hat\gamma}_k)^n-2
\sum_{\ell=1}^{j}\sum_{k=j+1}^{2j+1}({\gamma}_\ell/{\gamma}_k)^n \Bigg];
\nonumber\\
\fl
\ln\mathcal{H}=\!\sum_{n=1}^\infty \frac 1 {2n}\Bigg[
\sum_{\ell=1}^{j+1}\sum_{k=j+1}^{2j+1}({\hat\gamma}_\ell/{\gamma}_k)^n
+\!\sum_{\ell=j+2}^{2j+1}\sum_{k=1}^{j}({\gamma}_k/{\hat\gamma}_\ell)^n
\!-\!\sum_{\ell=1}^{j+1}\sum_{k=1}^{j}({\hat\gamma}_\ell{\gamma}_k)^n
\nonumber\\
-\sum_{\ell=j+2}^{2j+1}\sum_{k=j+1}^{2j+1}({\hat\gamma}_\ell{\gamma}_k)^{-n}\Bigg].
\ea
Consequently, using (\ref{mag}) and
\be
\bigg(\sum_{\ell=j'}^{k'}{\gamma}^n_\ell\bigg)^2=
\sum_{\ell=j'}^{k'}{\gamma}^{2n}_\ell+\sum_{j'\le\ell< k\le k'}\!2({\gamma}_\ell{\gamma}_k)^{n},
\ee
we find
\ba\fl
\mathcal{F}=\prod_{\ell=1}^{j}(1-\gamma^2_{\ell})^{1/4}
\prod_{\ell=1}^{j+1}(1-{\hat\gamma}^{2}_{\ell})^{1/4}
\prod_{\ell=j+2}^{2j+1}[(1-\gamma^{-2}_{\ell})(1-{\hat\gamma}^{-2}_{\ell})]^{1/4}
\!\!\prod_{1\le\ell< k\le j}\!(1-{\gamma}_{\ell}{\gamma}_{k})^{\halfs}
\nonumber\\
\fl
\hspace{10pt}\prod_{1\le\ell< k\le j+1}\!(1-{\hat\gamma}_{\ell}{\hat\gamma}_{k})^{\halfs}
\!\!\!\prod_{1+j\le\ell< k\le 2j+1}\!\!(1-({\gamma}_{\ell}{\gamma}_{k})^{-1})^{\halfs}
\!\!\!\prod_{2+j\le\ell< k\le 2j+1}\!\!(1-({\hat\gamma}_{\ell}{\hat\gamma}_{k})^{\!-1})^{\halfs}
\nonumber\\
\fl\hspace{20pt}
\prod_{\ell=1}^{j}\prod_{k=j+2}^{2j+1}(1-{\gamma}_{\ell}/{\hat\gamma}_{k})^{-\halfs}
\prod_{\ell=1}^{j+1}\prod_{k=j+2}^{2j+1}(1-{\hat\gamma}_{\ell}/{\hat\gamma}_{k})^{-\halfs};
\nonumber\\
\fl
\mathcal{H}=
\prod_{\ell=1}^{j}\prod_{k=1}^{j+1}(1-{\gamma}_{\ell}{\hat\gamma}_{k})^{\halfs}
\prod_{\ell=j+1}^{2j+1}\prod_{k=j+2}^{2j+1}(1-({\gamma}_{\ell}{\hat\gamma}_{k})^{\!-1})^{\halfs}
\nonumber\\
\fl\hspace{25pt}
\prod_{\ell=1}^{j}\prod_{k=j+1}^{2j+1}(1-{\gamma}_{\ell}/{\gamma}_{k})^{-\halfs}
\prod_{\ell=1}^{j+1}\prod_{k=j+1}^{2j+1}(1-{\hat\gamma}_{\ell}/{\gamma}_{k})^{-\halfs}.
\ea
The sum over $n$ can be carried out to obtain logarithmic functions.
Consequently, we find that (\ref{mag}) becomes
\ba
{M}^2=(1-\gamma^{-2}_{j+1})^{1/4}\mathcal{F}\mathcal{H}.
\label{mag1}\ea

\subsection{Spontaneous magnetization at the center of the string}

We substitute (\ref{Phip}) into (\ref{mgn}) and use (\ref{AB}) to obtain
\ba
g'_n=- g'_{-n}=\frac 1{2n}\left[-\sum_{\ell=1}^{j+1}{\hat\gamma}^n_\ell+
\sum_{\ell=j+2}^{2j+1}{\hat\gamma}^{-n}_\ell
+\sum_{\ell=1}^{j}\gamma^n_\ell-\sum_{\ell=j+1}^{2j+1}\gamma^{-n}_\ell\right],
\label{mgn2}\ea
such that
\ba\fl
\sum_{n=1}^\infty n g'_n g'_{-n}=-\frac 1 4\sum_{n=1}^\infty
\frac 1 n \left[-\sum_{\ell=1}^{j+1}{\hat\gamma}^n_\ell+\sum_{\ell
=j+2}^{2j+1}{\hat\gamma}^{-n}_\ell+\sum_{\ell=1}^{j}\gamma^n_\ell-
\sum_{\ell=j+1}^{2j+1}\gamma^{-n}_\ell\right]^2.
\ea
Comparing with (\ref{ngg}), and again carrying out the sum over $n$ , we find
that spontaneous magnetization at the center of the strings is
\ba
{M'}^2=(1-\gamma^{-2}_{j+1})^{1/4}\mathcal{F}/\mathcal{H}.
\label{mag1p}\ea

\section{Asymptotic behavior of the correlations function above \boldmath{$T_{\mathrm{c}}$}}

For $T>T_{\mathrm{c}}(1,m,n)$, the root $\gamma_{j+1}$ of $B(\theta)$ becomes smaller
than 1, and the spontaneous magnetization
is identically zero, and the kernel $\Phi(\theta)$ in (\ref{Phi}) (or $\Phi'(\theta)$ in (\ref{Phip}))
of the Wiener-Hopf sum equations has index $-1$, as can be seen
on page 209 and (1.7) on page 250 of McCoy and Wu's book \cite{MWbk}.
We follow the method described on pages 251--255 of \cite{MWbk}, to define
\ba
\Phi_1(\theta)=\Phi(\theta)\re^{\ri\theta}, \quad b_n=a_{n-1}=
\frac 1{2\pi}\int_{-\pi}^{\pi}\rd \theta\, \re^{-\ri n\theta}\,\Phi_1(\theta),\\
\Phi'_1(\theta)=\Phi'(\theta)\re^{\ri\theta}, \quad b'_n=a'_{n-1}=
\frac 1{2\pi}\int_{-\pi}^{\pi}\rd \theta\, \re^{-\ri n\theta}\,\Phi'_1(\theta).
\ea
We also let $R_r$ be the $r\times r$ Toeplitz determinant formed by $b_n$.
From the linear equations, (see (2.5) on page 251 in \cite{MWbk}),
\be
\sum_{m=0}^r b_{n-m} x_m^{(r)}=\delta_{n,0},\qquad
\sum_{m=0}^r b'_{n-m} {x'}_m^{(r)}=\delta_{n,0},
\ee
we find that the correlation above $T_{\mathrm{c}}$ is given as
\ba
\langle \sigma_{0,1}\sigma_{0,r+1}\rangle=(-1)^r R_{r+1}x_r^{(r)},
\label{corr}\\
\langle \sigma_{\bn,1}\sigma_{\bn,r+1}\rangle=(-1)^r R'_{r+1}{x'}_r^{(r)}.
\label{corrp}\ea
By solving the Wiener-Hopf equations as on pages 252--253 in \cite{MWbk},
we find as given in (2.27) on page 253 in \cite{MWbk} that
\ba\fl
x_r^{(r)}=\frac 1{2\pi i}\oint\rd \xi\, \xi^{r-1}\,\frac{P_1(\xi^{-1})}{Q_1(\xi)}, \quad
{x'}_r^{(r)}=\frac 1{2\pi i}\oint\rd \xi\, \xi^{r-1}\,\frac{P_1'(\xi^{-1})}{Q_1'(\xi)},
\quad \xi=\re^{\ri\theta},
\label{xrr}\ea
where
\ba
\Phi_1(\theta)=\sqrt\frac{\re^{2i\theta}\overline{A(\theta)}\,\overline {B(\theta)}}
{A(\theta)B(\theta)}=\frac 1{P_1(\xi)Q_1(\xi^{-1})},
\label{phi1}\\
\Phi'_1(\theta)=\sqrt\frac{\re^{2i\theta} {A(\theta)}\,\overline {B(\theta)}}
{\overline A(\theta)B(\theta)}=\frac 1{P'_1(\xi)Q'_1(\xi^{-1})}.
\label{phi1p}\ea
To calculate the integrals in (\ref{xrr}), we deform the contour of integration around
the branch cuts inside the unit circle. But here we have $2(2j+1)$ branch points,
as seen from (\ref{Phi}), (\ref{Phip}) and (\ref{AB}), instead of 2 as on page 254 of
\cite{MWbk}. The roots of $B(\theta)$ obtained by numerical
calculation in Maple are ordered as $\gamma_1\le\gamma_2\cdots\le \gamma_{2j+1}$,
so are the roots of $A(\theta)$, with
${\hat\gamma}_1\le{\hat\gamma}_2\cdots\le {\hat\gamma}_{2j+1}$.
We also find ${\hat\gamma}_n<{\gamma}_n$ for same $n$.
The integral becomes a sum of integrations around the branch cuts
from ${\hat\gamma}_n$ to $\gamma_n$ for $1\le n\le j+1$, and from
$1/\gamma_n$  to $1/{\hat\gamma}_n$ for $j+2\le n\le 2j+1$.

\subsection{Correlation for $T>T_{\mathrm{c}}$ at the center of the strip}

From (\ref{phi1}) and (\ref{AB}) for the correlation on the center of the strip, we find 
\ba
P_1(\xi)=\left[\prod_{\ell=j+2}^{2j+1}[(1-\gamma_\ell^{-1}\xi)(1-{\hat\gamma}_\ell^{-1}\xi)]
\Bigg/\prod_{\ell=1}^{j+1}[(1-\gamma_\ell\xi)(1-{\hat\gamma}_\ell\xi)]\right]^{\halfs},
\nonumber\\
Q_1(\xi)=\left[\prod_{\ell=1}^{j+1}[(1-\gamma_\ell\xi)(1-{\hat\gamma}_\ell\xi)]
\Bigg/\prod_{\ell=j+2}^{2j+1}[(1-\gamma_\ell^{-1}\xi)(1-{\hat\gamma}_\ell^{-1}\xi)]\right]^{\halfs}.
\label{PQ}\ea
So that (\ref{xrr}) becomes
\ba
x_r^{(r)}=\frac 1{\pi\ri}\left[\sum_{n=1}^{j+1} \int_{{\hat\gamma}_n}^{{\gamma}_n}\rd \xi\,
\xi^{r-1}\,\frac{P_1(\xi^{-1})}{Q_1(\xi)}
+\sum_{n=j+2}^{2j+1} \int_{1/{\gamma}_n}^{1/{\hat\gamma}_n}\rd \xi\,
\xi^{r-1}\,\frac{P_1(\xi^{-1})}{Q_1(\xi)}\right].
\label{xrrsum}\ea
If we let $\xi=\gamma_n\xi'$ for integrals in the first sum and $\xi=\xi'/{\hat\gamma}_n$
for the integrals in the second sum, and denote
$u_n={\hat\gamma}_n/{\gamma}_n$,  we may show that
\ba
\int_{{\hat\gamma}_n}^{{\gamma}_n}\rd \xi\, \xi^{r-1}\,
\frac{P_1(\xi^{-1})}{Q_1(\xi)}=\gamma_n^r
\int_{u_n}^{1}\rd \xi'\, \xi'^{r-1}\,\frac{P_1((\gamma_n\xi')^{-1})}{Q_1(\gamma_n\xi')},
\nonumber\\
\int_{1/{\gamma}_n}^{1/{\hat\gamma}_n}\rd \xi\, \xi^{r-1}\,
\frac{P_1(\xi^{-1})}{Q_1(\xi)}={\hat\gamma}_n^{-r}
\int_{u_n}^{1}\rd \xi'\, \xi'^{r-1}\,\frac{P_1({\hat\gamma}_n/\xi')}{Q_1(\xi'/{\hat\gamma}_n)}.
\label{intcuts}\ea
Since $\gamma_n<\gamma_{j+1}$ for $1\le n\le j$ and
${\hat\gamma}^{-1}_n<\gamma_{j+1}$ for $2+j\le n\le 2j+1$, in the
asymptotic limit $r>>1$, only one term is left, and it is
\ba
x_r^{(r)}\doteqdot\frac {\gamma^r_{j+1}}{\pi\ri} \int_{u_{j+1}}^1\rd \xi\, \xi^{r-1}\,\frac{P_1((\gamma_{j+1}\xi)^{-1})}{Q_1(\gamma_{j+1}\xi)}.
\label{xrr1}\ea
We now use (\ref{PQ}) to expand asymptotically
\ba\fl
P_1((\gamma_{j+1}\xi)^{-1})\!\doteqdot\frac\xi{\ri[(1-\xi)(\xi-u_{j+1})]^{\halfs}}
\frac{\prod_{\ell=j+2}^{2j+1}[(1-(\gamma_{j+1}\gamma_\ell)^{-1})
(1-(\gamma_{j+1}{\hat\gamma}_\ell)^{-1})]^{\halfs}}
{\prod_{\ell=1}^{j}[(1-\gamma_\ell/\gamma_{j+1})
(1-{\hat\gamma}_\ell/\gamma_{j+1})]^{\halfs}}+\cdots,
\nonumber\\
\fl
Q_1(\gamma_{j+1}\xi)\!\doteqdot\frac{\prod_{\ell=1}^{j+1}
[(1-\gamma_\ell\gamma_{j+1})(1-{\hat\gamma}_\ell\gamma_{j+1})]^{\halfs}}
{\prod_{\ell=j+2}^{2j+1}[(1-\gamma_{j+1}/\gamma_\ell)
(1-\gamma_{j+1}/{\hat\gamma}_\ell)]^{\halfs}}+\mathrm{O}(1-\xi).
\label{PQ1}\ea
Consequently, (\ref{xrr1}) becomes
\ba\fl
x_r^{(r)}\doteqdot-\frac {\gamma^r_{j+1}\mathcal{U}}{\pi } \int_{u_{j+1}}^1 \,
\frac{\rd \xi\,\xi^{r}}{[(1-\xi)(\xi-u_{j+1})]^{\halfs}}
=-\frac {\gamma^r_{j+1}\mathcal{U}}{\pi }  \int_0^1\rd y\,
\frac{[1-(1-u_{j+1})y]^{r}}{[y(1-y)]^{\halfs}},
\label{xrr2}\ea
where we have changed the variable $\xi=1-(1-u_{j+1})y$ and 
\ba\fl
\hspace{45pt}
\mathcal{U}=\prod_{\ell=j+2}^{2j+1}[(1-(\gamma_{j+1}\gamma_\ell)^{-1})
(1-(\gamma_{j+1}{\hat\gamma}_\ell)^{-1})
(1-\gamma_{j+1}/\gamma_\ell)(1-\gamma_{j+1}/{\hat\gamma}_\ell)]^{\halfs}
\nonumber\\
\prod_{\ell=1}^{j}[(1-\gamma_\ell/\gamma_{j+1})(1-{\hat\gamma}_\ell/\gamma_{j+1})]^{-\halfs}
\prod_{\ell=1}^{j+1}[(1-\gamma_\ell\gamma_{j+1})(1-{\hat\gamma}_\ell\gamma_{j+1})]^{-\halfs}.
\ea
Using
\be
(1-x)^{r}=\sum_{\ell=0}^r\frac{(-r)_\ell }{\ell!}x^\ell,
\ee
and \cite[GR8.380]{GR}
\be
\int_0^1\rd y\,y^{\alpha-1}(1-y)^{\beta-1}=B(\alpha,\beta)=
\frac{\Gamma(\alpha)\Gamma(\beta)}{\Gamma(\alpha+\beta)},
\ee
we find 
\ba
x_r^{(r)}\doteqdot -\mathcal{U}\gamma^r_{j+1} \,\,{}_{2} F_1
\left[{-r,\half \atop{\phantom {\omega}1}};1-u_{j+1}\right],\quad
u_{j+1}={\hat\gamma}_{j+1}/{\gamma}_{j+1}.
\label{xrr3}
\ea
We again use Szeg\H o's theorem to calculate the Toeplitz determinant,
\ba \lim_{r\to\infty}(-1)^{r} R_{r}=
\mathcal{F}\mathcal{H}(1-{\hat\gamma}_{j+1}/\gamma_{j+1})^{1/2}
(1-\gamma^2_{j+1})^{-1/4}\mathcal{U}^{-1},
\label{Rr}\ea
so that 
\ba\fl
\langle\sigma_{0,1}\sigma_{0,r+1}\rangle=
-\lim_{r\to\infty}(-1)^{r+1} R_{r+1}x_r^{(r)}\doteqdot
\frac {\gamma^r_{j+1}\mathcal{F}\mathcal{H}(1-u_{j+1})^{\halfs}}{(1-\gamma^2_{j+1})^{1/4}}
\,{}_{2} F_1\left[{-r,\half \atop{\phantom {\omega}1}};1\!-\!u_{j+1}\right].
\label{corr1gto}\ea

Using \cite[9.131.2]{GR}, we have
\ba
{}_{2} F_1\left[{-r,\half \atop{\phantom {\omega}1}};1\!-\!u_{j+1}\right]
=\frac {\Gamma(r\!+\!1/2)}{\Gamma(1/2)\Gamma(r+1)}\,\,
{}_{2} F_1\left[{-r,\half \atop{\phantom {1}\half-r}};u_{j+1}\right]. 
\label{xrr3o}\ea
In the limit $r\gg1$, we find
\be
 \frac {\Gamma(r\!+\!1/2)}{\Gamma(r+1)}\to r^{-\halfs},\qquad
 {}_{2} F_1\left[{-r,\half \atop{\phantom {1}\half-r}};u\right] \to (1-u)^{-\halfs}.
 \label{hypg}\ee 
Whenever (\ref{hypg}) is valid, we may substitute this approximation into (\ref{corr1gto})
and find that the leading term in the spin-spin correlation function above $T_{\mathrm{c}}$
behaves as
\ba
\langle \sigma_{0,1}\sigma_{0,r+1}\rangle\doteqdot
\frac {\gamma^r_{j+1}\mathcal{F}\mathcal{H}}{(1-\gamma^2_{j+1})^{1/4}\sqrt{\pi r}}+\cdots.
\label{corr1gt}\ea
It is interesting to examine this equation using the spontaneous magnetization
given by (\ref{mag1}): We find it has the same behavior as in the regular Ising model
by comparing it with (2.45) on page 256 and with (2.8) on page 245 of \cite{MWbk}.  

It was shown in our earlier paper \cite{HJPih}, that for strings of length $n\ge5$
a rounded peak in the specific heat shows up above $T_{\mathrm{c}}(1,m,n)$
due to the finite strip width $m=2j$. Near $T_{max}$
(or $z_{max}=\tanh(J/T_{max}k_{\mathrm{B}})$) of the maximum in the specific heat,
we find $\gamma_{j+1}<1$, smaller than its value near $T_{\mathrm{c}}(1,m,n)$,
which is near 1. For (\ref{hypg}) to be valid, we must have $r$ extremely large,
such that $\gamma^r_{j+1}$ is ignorable, not like the case with $T$ near
$T_{\mathrm{c}}(1,m,n)$. When $r$ is not so large, (\ref{corr1gto}) must be used,
which exhibits the one-dimensional Ising behavior, just like the specific heat.

When we look at the case with $n=8$ and $j=6$, we find that the rounded peak
of the specific heat is at its maximum at the temperature $T_{max}$ corresponding
to $z_{max}=0.4730$. At that value, we have $\gamma_{7}=0.990166387243$,
and $u_{7}=0.995068080479$. To have (\ref{hypg}) valid, we need to have
$r\gg12000$. But $\gamma_{7}^r\approx0$, which is different from critical behavior,
(as $T\to T_{\mathrm{c}}$, $\gamma_{7}\to1$).
For $r<2000$, so that $\gamma_{7}^r\ge10^{-10}$, we have to use (\ref{xrr3})
for the correlation function, so that
\ba
\langle \sigma_{0,1}\sigma_{0,r+1}\rangle\doteqdot\mathcal{G}_0\gamma^r_{7} \,\,
{}_{2} F_1\left[{-r,\half \atop{\phantom {\omega}1}};1\!-\!u_{7}\right], \quad
\mathcal{G}_0=1.31179304205778.
\label{corr1gtm}\ea
This behaves as in the one-dimensional Ising model. We have also checked
this for different values of $j$, reaching the same conclusion.

\subsection{Correlation for $T>T_{\mathrm{c}}$ at the center of the string}

Similarly, for the correlation on the center of the string, we find from (\ref{phi1})
and (\ref{AB}) that
\ba
P'_1(\xi)=\left[\prod_{\ell=1}^{j+1}\frac{1-{\hat\gamma}_\ell\xi}{1-\gamma_\ell\xi}\,
\prod_{\ell=j+2}^{2j+1}
\frac{1-\gamma_\ell^{-1}\xi}{1-{\hat\gamma}_\ell^{-1}\xi}\right]^{\halfs},
\nonumber\\
Q'_1(\xi)=\left[\prod_{\ell=1}^{j+1}\frac{1-\gamma_\ell\xi}{1-{\hat\gamma}_\ell\xi}\,
\prod_{\ell=j+2}^{2j+1}\frac{1-{\hat\gamma}_\ell^{-1}\xi}{1-\gamma_\ell^{-1}\xi}\right]^{\halfs}.
\label{PQp}\ea
As many of the steps are similar, we will be more brief here. Again deforming the contour
of integration to be around the $2j+1$ branch cuts, we find that, as in (\ref{xrr1}), only one
term is left in the asymptotic limit $r\gg1$. We further write, as in (\ref{PQ1}),
\ba
\frac{P_1'((\gamma_{j+1}\xi)^{-1})}{Q_1'(\gamma_{j+1}\xi)}\doteqdot
\left[\frac{\xi-u_{j+1}}{\xi-1}\right]^{\halfs}\mathcal{U}'+\mathrm{O}(1-\xi),
\label{PQp2}\ea
in which $u_{j+1}={\hat\gamma}_{j+1}/\gamma_{j+1}$ and
\ba
\mathcal{U}'=&\prod_{\ell=1}^{j}\left[\frac{1-{\hat\gamma}_\ell/\gamma_{j+1}}
{1-\gamma_\ell/\gamma_{j+1})}\right]^{\halfs}
\prod_{\ell=1}^{j+1}\left[\frac{1-{\hat\gamma}_\ell\gamma_{j+1}}
{1-\gamma_\ell\gamma_{j+1}}\right]^{\halfs}
\nonumber\\
&\prod_{\ell=j+2}^{2j+1}\left[\frac{(1-(\gamma_{j+1}\gamma_\ell)^{-1})
(1-\gamma_{j+1}/\gamma_\ell)}{(1-(\gamma_{j+1}{\hat\gamma}_\ell)^{-1})
(1-\gamma_{j+1}/{\hat\gamma}_\ell)}\right]^{\halfs}
\ea
such that replacing $P_1$ and $Q_1$ in (\ref{xrr1}) by their primed versions, we have
\ba
{x'}_r^{(r)}&\doteqdot-\frac {\gamma^r_{j+1}}{\pi }\mathcal{U}'(1-u_{j+1})
\int_0^1\rd y\,\left[\frac{1-y}y\right]^{\halfs}[1-(1-u_{j+1})y]^{r-1}
\nonumber\\
&=-\half {\gamma^r_{j+1}} \,\mathcal{U}'(1-u_{j+1})\,
{}_{2} F_1\left[{1-r,\half \atop{\phantom {\omega}2}};1\!-\!u_{j+1}\right]
\nonumber\\
&=-\frac {\gamma^r_{j+1}\Gamma(r\!+\!1/2)}{\sqrt{\pi}\Gamma(r+1)} \,\mathcal{U}'(1-u_{j+1})\,
{}_{2} F_1\left[{1-r,\half \atop{\phantom {1}\half-r}};u_{j+1}\right]
\nonumber\\
&\approx -\frac{\gamma^r_{j+1}}{\sqrt{\pi r}}\,\mathcal{U}'(1-u_{j+1})^{\halfs}.
\label{xrr2p}\ea
It is also easy to show 
\ba \lim_{r\to\infty}(-1)^{r} R'_{r}=[\mathcal{F}/\mathcal{H}]
(1-{\hat\gamma}_{j+1}/\gamma_{j+1})^{-1/2}(1-\gamma^2_{j+1})^{-1/4}{\mathcal{U}'}^{-1}.
\ea
Consequently, we find
\ba
\langle \sigma_{\bn,1}\sigma_{\bn,r+1}\rangle=
-\lim_{r\to\infty}(-1)^{r+1} R'_{r+1}{x'}_r^{(r)}\doteqdot\frac {\gamma^r_{j+1}
[\mathcal{F}/\mathcal{H}]}{(1-\gamma^2_{j+1})^{1/4}\sqrt{\pi r}}+\cdots.
\label{corr1gtp}\ea

\section{Asymptotic behavior of the correlations function for \boldmath{$T<T_{\mathrm{c}}$}}

For $T<T_{\mathrm{c}}(1,m,n)$, the root $\gamma_{j+1}$ of $B(\theta)$ is greater than 1.
The spontaneous magnetization is non-zero and given by (\ref{mag1}) and (\ref{mag1p}).
We again follow the method described on pages 257--258 in \cite{MWbk}, to calculate
the asymptotic behavior for $T<T_{\mathrm{c}}$. 
Rewrite (3.4)--(3.6) on page 257 of \cite{MWbk} for $\Phi(\theta)$ and $\Phi'(\theta)$ as
\ba
\Phi(\theta)=\sqrt\frac{\overline {A(\theta)}\,\overline {B(\theta)}}{A(\theta)B(\theta)}=\frac 1{P(\xi)Q(\xi^{-1})},
\label{phipq}\\
\Phi'(\theta)=\sqrt\frac{{A(\theta)}\,\overline {B(\theta)}}{\overline A(\theta)B(\theta)}=\frac 1{P'(\xi)Q'(\xi^{-1})},
\label{phipqp}\ea
which are similar to (\ref{phi1}) and (\ref{phi1p}). We can then use (3.14) on
page 258 of \cite{MWbk} to calculate the asymptotic behavior of the correlations
function for $T<T_{\mathrm{c}}$.

\subsection{Correlation for $T<T_{\mathrm{c}}$ at the center of the strip}

For $T<T_{\mathrm{c}}$, the asymptotic behavior of the correlation at the center
of a strip is therefore given by (3.14) on page 258 of \cite{MWbk} as
\ba\fl
M^{-2}\langle \sigma_{0,1}\sigma_{0,r+1}\rangle\doteqdot 1+\frac 1{4\pi^2}
\oint\rd \xi\, \xi^{r}\,\frac{Q(\xi^{-1})}{P(\xi)}
\oint\, \frac{\rd \xi'(1/\xi')^{r}}{(\xi'-\xi)^2}\,\frac{P(\xi')}{Q({\xi'}^{-1})}\doteqdot1+\mathcal{S}_r.
\label{corrlt}\ea
The integration over $\xi$ is deformed to be around the branch cuts inside the unit circle;
while the integration over $\xi'$ is around the branch cuts outside the unit circle.
When $T<T_{\mathrm{c}}$, we find
${\hat\gamma}_1\le{\gamma}_1\le{\hat\gamma}_2\le{\gamma}_2\cdots
\le {\hat\gamma}_{j+1}\le {\gamma}^{-1}_{j+1}<1$,
and ${\gamma}^{-1}_{2j+1}\le{\hat\gamma}^{-1}_{2j+1}\cdots
\le{\gamma}^{-1}_{j+2}\le{\hat\gamma}^{-1}_{j+2}<1$.
These are the branch points inside the unit circle. The branch points outside
the unit circle are their inverses. Thus, we find
\ba\fl
\mathcal{S}_r=&\frac 1{\pi^2}\Bigg\{\Bigg[\sum_{\ell=1}^j
\int_{{\hat\gamma}_\ell}^{{\gamma}_\ell}\rd \xi\, \xi^{r}\,\frac{Q(1/\xi)}{P(\xi)}
+\int_{{\hat\gamma}_{j+1}}^{{\gamma}^{-1}_{j+1}}\rd \xi\, \xi^{r}\,\frac{Q(1/\xi)}{P(\xi)}
+\sum_{\ell=j+2}^{2j+1}\int_{{\gamma}^{-1}_\ell}^{{\hat\gamma}^{-1}_\ell}\rd \xi\,
\xi^{r}\,\frac{Q(1/\xi)}{P(\xi)}\Bigg]
\nonumber\\
\fl&\times
\Bigg[\sum_{\ell=1}^j\!\int_{{\gamma}^{-1}_\ell}^{{\hat\gamma}^{-1}_\ell}
\frac{\rd \xi'(1/\xi')^{r}}{(\xi'-\xi)^2}\frac{P(\xi')}{Q(1/{\xi'})}
+\int_{{\gamma}_{j+1}}^{{\hat\gamma}^{-1}_{j+1}}\frac{\rd \xi'(1/\xi')^{r}}
{(\xi'-\xi)^2}\frac{P(\xi')}{Q(1/{\xi'})}
\nonumber\\
\fl&\qquad
+\sum_{\ell=j+2}^{2j+1}\int_{{\gamma}_\ell}^{{\hat\gamma}_\ell}
\frac{\rd \xi'(1/\xi')^{r}}{(\xi'-\xi)^2}\frac{P(\xi')}{Q(1/{\xi'})}\Bigg]\Bigg\}.
\ea
Similar to (\ref{intcuts}), we may estimate these integrals and find in the
limit $r\gg1$, only one term is needed, namely
\ba
\mathcal{S}_r\doteqdot\frac 1{\pi^2}\int_{{\hat\gamma}_{j+1}}^{{\gamma}^{-1}_{j+1}}\rd \xi\,
\xi^{r}\,\frac{Q(1/\xi)}{P(\xi)}\int_{{\gamma}_{j+1}}^{{\hat\gamma}^{-1}_{j+1}}
\frac{\rd \xi'(1/\xi')^{r}}{(\xi'-\xi)^2}\frac{P(\xi')}{Q(1/{\xi'})}.
\label{Sr1}\ea
Let $\xi=\xi_1/{\gamma}_{j+1}$ and $\xi'=\gamma_{j+1}/\xi_2$, then the above equation becomes
\ba\fl
\mathcal{S}_r\doteqdot\frac {\gamma^{-2r}_{j+1}}{\pi^2}\int_{v_{j+1}}^1\rd \xi_1\, \xi_1^{r}\,
\frac{Q(\gamma_{j+1}/\xi_1)}{P(\xi_1/{\gamma}_{j+1})}
\int_{v_{j+1}}^1\frac{\rd \xi_2\,\xi_2^{r}}{(\gamma_{j+1}-\xi_1\xi_2/\gamma_{j+1})^2}\,
\frac{P(\gamma_{j+1}/\xi_2)}{Q(\xi_2/\gamma_{j+1})},
\label{Sr2}\ea
in which $v_{j+1}=\gamma_{j+1}{\hat\gamma}_{j+1}$.
Similar to (\ref{PQ1}) or (\ref{PQp2}), we pull out the singular terms of the integrands,
and expand what remains around $\xi_1=1$ and $\xi_2=1$ and find
\ba\fl
\frac1{(\gamma_{j+1}-\xi_1\xi_2/\gamma_{j+1})^2}\doteqdot
\frac1{(\gamma_{j+1}-1/\gamma_{j+1})^2}
+\mathrm{O}(1-\xi_1,1-\xi_2),\label{gg12}\\
\fl
\frac{Q(\gamma_{j+1}/\xi_1)P(\gamma_{j+1}/\xi_2)}
{P(\xi_1/{\gamma}_{j+1})Q(\xi_2/\gamma_{j+1})}\doteqdot\left[\frac{(\xi_1-v_{j+1})(\xi_2-1)}
{(\xi_1-1)(\xi_2-v_{j+1})}\right]^{\halfs}+\mathrm{O}(1-\xi_1,1-\xi_2).
\label{QPPQ}\ea
Substituting the above equations into (\ref{Sr2}), we find
\ba
\mathcal{S}_r\doteqdot\frac {\gamma^{-2r}_{j+1}\,I_1(v_{j+1})I_2(v_{j+1})}{\pi^2(\gamma_{j+1}-1/\gamma_{j+1})^2},
\nonumber\\
I_1(v)=\!\int_{v}^1\!\rd \xi\,\xi^{r}\Bigg[\frac{ \xi-v}{ 1-\xi}\Bigg]^{\halfs},\,
I_2(v)=\!\int_{v}^1\!\rd \xi\,\xi^{r}\Bigg[\frac{ 1-\xi}{ \xi-v}\Bigg]^{\halfs}.
\label{Sr3}\ea
Following the steps described from (\ref{xrr2}) to (\ref{hypg}), we find
\ba
I_1(v)=\half\pi(1-v)\,{}_{2} F_1\left[{-r,\half \atop{\phantom{\omega}2}};1\!-\!v\right]
\nonumber\\
\hspace{28pt}
=\sqrt{\pi}(1-v)\frac{\Gamma(r\!+\!1/2)}{\Gamma(r+1)}\,
{}_{2} F_1\left[{-r,\half \atop{-\half-r}};v\right]
\doteqdot\sqrt{\frac{\pi}{r}}(1-v)^{\halfs},\\
I_2(v)=\half\pi(1-v)\,{}_{2} F_1\left[{-r,3/2 \atop{\phantom {\omega}2}};1\!-\!v\right]
\nonumber\\
\hspace{28pt}
=\half\sqrt{\pi}(1-v)\frac{\Gamma(r\!+\!1/2)}{\Gamma(r+2)}\,
{}_{2} F_1\left[{-r,3/2\atop{\half-r}};v\right]
\doteqdot\half\sqrt{\frac{\pi}{r^3}}(1-v)^{-\halfs},
\ea
so that
\ba
\mathcal{S}_r\doteqdot\frac {\gamma^{-2r}_{j+1}}{2\pi\,r^2(\gamma_{j+1}-1/\gamma_{j+1})^2}.\label{Sr4}\ea
The asymptotic behavior of the correlation for $T<T_{\mathrm{c}}$ at the center
of the strip is seen from (\ref{corrlt}) to be
\ba
\langle \sigma_{0,1}\sigma_{0,r+1}\rangle\doteqdot M^{2} \left[1+\frac {\gamma^{-2r}_{j+1}}{2\pi\,r^2(\gamma_{j+1}-1/\gamma_{j+1})^2}\right].
\label{corr1lt}\ea

\subsection{Correlation for $T<T_{\mathrm{c}}$ at the center of the string}

The asymptotic behavior of the correlation at the center of the string
for $T<T_{\mathrm{c}}$ is similar to (\ref{corrlt}), 
 \ba\fl
{M'}^{-2}\langle \sigma_{\bn,1}\sigma_{\bn,r+1}\rangle\doteqdot 1+
\frac 1{4\pi^2}\oint\rd \xi\, \xi^{r}\,\frac{Q'(\xi^{-1})}{P'(\xi)}
\oint\, \frac{\rd \xi'(1/\xi')^{r}}{(\xi'-\xi)^2}\,\frac{P'(\xi')}{Q'({\xi'}^{-1})}
\doteqdot 1+\mathcal{S}'_r,
\label{corrltp}\ea
in which $P'(\xi)$ and $Q'(\xi)$ are given by (\ref{phipqp}). The same as in (\ref{Sr1}),
we find that in the limit $r\gg1$, only one term is needed, so that
\ba\fl
\mathcal{S}'_r\doteqdot\frac 1{\pi^2}
\int_{{\hat\gamma}_{j+1}}^{{\gamma}^{-1}_{j+1}}\rd \xi\, \xi^{r}\,\frac{Q'(1/\xi)}{P'(\xi)}
\int_{{\gamma}_{j+1}}^{{\hat\gamma}^{-1}_{j+1}}\frac{\rd \xi'(1/\xi')^{r}}{(\xi'-\xi)^2}
\frac{P'(\xi')}{Q'(1/{\xi'})}.
\label{Sr1p}\\\fl\hspace{15pt}
\doteqdot\frac {\gamma^{-2r}_{j+1}}{\pi^2}\int_{v_{j+1}}^1\rd \xi_1\, \xi_1^{r}\,
\frac{Q'(\gamma_{j+1}/\xi_1)}{P'(\xi_1/{\gamma}_{j+1})}
\int_{v_{j+1}}^1\frac{\rd \xi_2\,\xi_2^{r}}{(\gamma_{j+1}-\xi_1\xi_2/\gamma_{j+1})^2}\,\frac{P'(\gamma_{j+1}/\xi_2)}{Q'(\xi_2/\gamma_{j+1})},
\label{Sr2p}\ea
when the substitution used in (\ref{Sr2}) is again used.  
Just as in (\ref{QPPQ}), we may write
\ba
\fl
\frac{Q'(\gamma_{j+1}/\xi_1)P'(\gamma_{j+1}/\xi_2)}
{P'(\xi_1/{\gamma}_{j+1})Q'(\xi_2/\gamma_{j+1})}\doteqdot\left[\frac{\xi_1(\xi_2-v_{j+1})(1-\xi_2)}
{\xi_2(\xi_1-v_{j+1})(1-\xi_1)}\right]^{\halfs}+\mathrm{O}(1-\xi_1,1-\xi_2).
\label{QPPQp}\ea
Using (\ref{gg12}) and (\ref{QPPQp}) in (\ref{Sr2p}), we find
\ba
\mathcal{S}'_r\doteqdot\frac {\gamma^{-2r}_{j+1}\,I'_1(v_{j+1})I'_2(v_{j+1})}{\pi^2(\gamma_{j+1}-1/\gamma_{j+1})^2},
\nonumber\\
I'_1(v)=\!\int_{v}^1\!\frac{\rd \xi\,\xi^{r+1}}{[ (\xi-v)( 1-\xi)]^{\halfs}},\quad
I'_2(v)=\!\int_{v}^1\!\rd \xi\,\xi^{r-1}[( 1-\xi)(\xi-v)]^{\halfs}.
\label{Sr3p}\ea
Again, we follow the steps given from (\ref{xrr2}) to (\ref{hypg}) to find
\ba
I'_1(v)&=\pi\,{}_{2} F_1\left[{-1-r,\half \atop{\phantom{\omega}1}};1\!-\!v\right]
\nonumber\\
&=\sqrt{\pi}\frac{\Gamma(r\!+\!3/2)}{\Gamma(r+2)}\,{}_{2} F_1\left[{-1-r,\half \atop{-\half-r}};v\right]
\doteqdot\sqrt{\frac{\pi}{r}}(1-v)^{-\halfs},\\
I'_2(v)&=\frac {\pi(1-v)^2} 8\,{}_{2} F_1\left[{1-r,3/2 \atop{\phantom{\omega}3}};1\!-\!v\right]
\nonumber\\
&=\frac{\sqrt{\pi}(1-v)^2\Gamma(r\!+\!1/2)}{2\Gamma(r+2)}\,{}_{2} F_1\left[{1-r,3/2 \atop{\half-r}};v\right]
\doteqdot\half\sqrt{\frac{\pi}{r^3}}(1-v)^{\halfs}.
\ea
Consequently, we find from (\ref{Sr3p}) 
\ba
\mathcal{S}'_r\doteqdot\frac {\gamma^{-2r}_{j+1}}{2\pi\,r^2(\gamma_{j+1}-1/\gamma_{j+1})^2},\label{Sr4p}\ea
so that
 \ba
 \langle \sigma_{\bn,1}\sigma_{\bn,r+1}\rangle\doteqdot {M'}^{2}\left[1+\frac {\gamma^{-2r}_{j+1}}{2\pi\,r^2(\gamma_{j+1}-1/\gamma_{j+1})^2}\right],
\label{corr1ltp} \ea
which has the usual two-dimensional Ising behavior.

\section{Correlation function for the central row of a finite strip}

In the limit $n\to\infty$, we have shown in subsection 3.3 that the model is equivalent
to independent strips of width $m$ and the generating function for the spins in the
central row of a strip is given by (\ref{Phiz0}). It behaves as in the one-dimensional
Ising model, with its critical temperature at $T=0$ (or $z=1$). Above this critical temperature,
we can use (\ref{corr}) to calculate its asymptotic behavior. (We tested this formula for $j=0$,
which is a one-dimensional spin chain, and got the exact result).
As in (\ref{Rr}), we again use Szeg\H o's theorem given in (\ref{mag}) and (\ref{mgn})
to calculate the Toeplitz determinant of $(-1)^rR_r$ whose generating function is given
in (\ref{Phiz0}).

In the limit $m=2j\to \infty$, it becomes the two-dimensional Ising model. We drop the
term $\alpha^{-j}$ in the first equation of (\ref{ABz}), and find from (\ref{Phiz0})
\be
\Phi(\theta)=\Phi_1(\theta)\to\Omega^{1/2}.
\label{phi1IS}\ee
which is the generating function of the regular 2-d Ising, as it should.

For finite $m$, we substitute the first equation in (\ref{AB}) into (\ref{Phiz0}) and find
\be
\Phi_1(\theta)=-\prod_{\ell=1}^{j+1}
\left[\frac{1-\gamma_\ell\re^{\ri\theta}}{1-\gamma_\ell\re^{-\ri\theta}}\right]
\prod_{\ell=j+2}^{2j+1}\left[\frac{1-\gamma^{-1}_\ell\re^{-\ri\theta}}
{1-\gamma^{-1}_\ell\re^{\ri\theta}}\right]
=\frac 1{P_1(\xi)Q_1(\xi^{-1})},
\label{phi1fn}\ee
Consequently, using Szeg\H o's theorem, we calculate the Toeplitz determinant and find 
\ba
\lim_{r\to\infty}(-1)^{r} R_{r}={\hat{\mathcal{F}}}_0/{\hat{\mathcal{U}}}_0,
\label{rrfn}\ea
 in which
 \ba
 {\hat{\mathcal{F}}}_0=&\prod_{\ell=1}^{j}
 \left[\frac{1-\gamma_\ell\gamma_{j+1}}{1-\gamma_\ell/\gamma_{j+1}}\right]
 \prod_{\ell=j+2}^{2j+1}\left[\frac{1-\gamma^{-1}_\ell/\gamma_{j+1}}
 {1-\gamma^{-1}_\ell\gamma_{j+1}}\right]
 \nonumber\\
&\times \frac{\prod_{\ell=1}^{j}\prod_{k=1}^{j}(1-\gamma_\ell\gamma_k)
 \prod_{\ell=j+2}^{2j+1}\prod_{k=j+2}^{2j+1}(1-\gamma^{-1}_\ell/\gamma_k)}
 {\prod_{\ell=1}^{j}\prod_{k=j+2}^{2j+1}(1-\gamma_\ell/\gamma_k)},
\ea 
and
 \ba
 {\hat{\mathcal{U}}}^{-1}_0=\frac{(1-\gamma^2_{j+1})
 \prod_{\ell=1}^{j}[(1-\gamma_\ell\gamma_{j+1})(1-\gamma_\ell/\gamma_{j+1})]}
{\prod_{\ell=j+2}^{2j+1}[(1-\gamma^{-1}_\ell\gamma_{j+1})(1-\gamma^{-1}_\ell/\gamma_{j+1})]}.
\ea 
From(\ref{xrr}), and (\ref{phi1fn}) we have
\ba\fl
x_r^{(r)}=\frac 1{2\pi\ri}\oint\rd \xi\, \xi^{r-1}\,\frac{P_1(\xi^{-1})}{Q_1(\xi)}=
\frac {-1}{2\pi\ri}\oint\rd \xi\, \xi^{r}\,
\frac{\prod_{\ell=j+2}^{2j+1}[(\xi-\gamma^{-1}_\ell)(1-\gamma^{-1}_\ell\xi)]}
{\prod_{\ell=1}^{j+1}[(\xi-\gamma_\ell)(1-\gamma_\ell\xi)]}.
\ea
This integral can be calculated by computing the residues at the poles
inside the unit circle. They are the $j+1$ roots $\gamma_k$ for $1\le k\le j+1$.
Since $\gamma_1\le\gamma_2\cdots\le\gamma_{j+1}$, we find for $r\gg1$,
we only need one term, namely the residue at $\gamma_{j+1}$, which is
\ba
x_r^{(r)}\approx-\frac{\gamma_{j+1}^r \prod_{\ell=j+2}^{2j+1}
[(1-\gamma^{-1}_\ell\gamma_{j+1})(1-\gamma^{-1}_\ell/\gamma_{j+1})]}
{(1-\gamma^2_{j+1})\prod_{\ell=1}^{j}[(1-\gamma_\ell\gamma_{j+1})
(1-\gamma_\ell/\gamma_{j+1})]}
\approx-\gamma_{j+1}^r {\hat{\mathcal{U}}}_0.
\label{xrrfn}\ea
Consequently, using (\ref{rrfn}) and (\ref{xrrfn}), we find that the correlation function
for the central row of a finite strip is
\ba
\langle \sigma_{0,1}\sigma_{0,r+1}\rangle=
\lim_{r\to\infty}(-1)^{r+1} R_{r+1}(-x_r^{(r)})\doteqdot \gamma^r_{j+1}{\hat{\mathcal{F}}}_0.
\label{corr1fn}\ea
We now examine the exponential decaying factor of the correlation function
in the above equation. For $n$ finite, $\gamma_{j+1}=1$ at the critical temperature
$T_{\mathrm{c}}(1,m,n)$, $\gamma_{j+1}>1$ below the critical temperature,
and $\gamma_{j+1}<1$ for $T>T_{\mathrm{c}}(1,m,n)$. But in the limit
$n\to\infty$, we have $T_{\mathrm{c}}(1,m,\infty)=0$, so that $\gamma_{j+1}=1$
at $T=0$ ($z=1$). We next plot $\gamma_{j+1}$ as a function of $z$ for different $m=2j$.

\begin{figure}[hbt]
\begin{center}
\includegraphics[width=0.75\hsize]{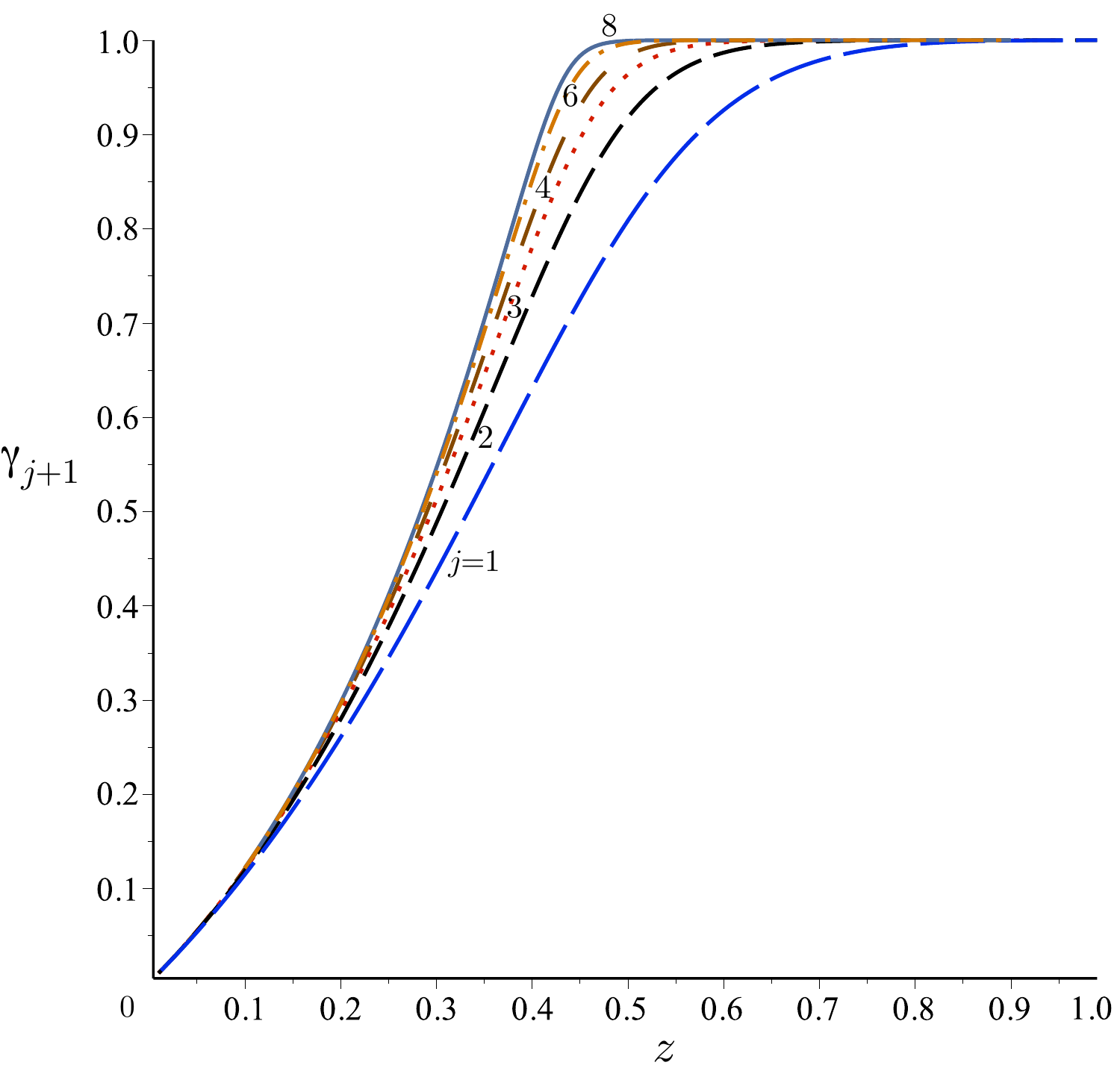}
\caption{(Color online) The root $\gamma_{j+1}$ in (\ref{corr1fn}) is plotted for
$m=2j=2,4,6,8,12,16,$, as a function of $z$. For $z=0$ ($T=\infty$), we have
$\gamma_{j+1}=0$, and for $z=1$ ($T=0$), we find $\gamma_{j+1}=1$. As shown
in the figure, as $z$ increases from zero, $\gamma_{j+1}$ increases and approaches 1.
In fact, $\gamma_{j+1}\approx1$ for a large region of $z$ near $z=1$. This means
that at sufficient low temperature, the exponential decay term is almost irrelevant,
and the correlation functions are almost a constant.  As $m$ increases, this region
of $z$, with $\gamma_{j+1}\approx1$, becomes larger.}
\end{center}
\label{fig:1}
\end{figure}

As shown in the figure, $\gamma_{j+1}$ is an increasing function of $z$ whose
maximum is 1 at $z=1$. It is also an increasing function of $m=2j$, (these
statements are observed but not rigorously proven). For example, we find that,
for $z = 0.75$, the exponential decay terms are given as
$\gamma_{j+1}= .9996300699194$, $.9999865761094$, $.9999995129623$,
$.9999999993589$, $.999999999999$ for $j= 2, 3, 4, 6, 8$. We can easily deduce
that as $m=2j$ increases, $\gamma_{j+1}$ approaches 1, so that the correlation
function behaves almost like a constant. We have shown in (\ref{phi1IS}), that
in the limit $m\to\infty$ the generating function becomes that of the regular Ising model,
whose correlation function for $T<T_{\mathrm{c}}$ (or $z>\sqrt{2}-1$ when $J=J'$) tends
to the constant value $M_0^2$. Thus these results make sense.

\section{Summary}

We have found that the spin correlation function in the central row of a strip in our model
pictured in figure 1 is given as Toeplitz determinant (\ref{Toeplitz}), whose
generating function is given in (\ref{Phi}). Likewise the correlation function in the
central row of a string row is also a Toeplitz determinant whose generating function
is given in (\ref{Phip}).

The spontaneous magnetization is different from row to row. At the center of a strip, it is given by (\ref{mag1}), and at the center of a string row it is (\ref{mag1p}). 

For $T>T_{\mathrm{c}}(1,m,n)$, but very near the critical temperature, the correlation
functions behave just like the two-dimensional regular Ising model and are given
by (\ref{corr1gt}) for the central rows of strips and by (\ref{corr1gtp}) for centers of
string rows. Near $T_{max}$, where the specific heat has its rounded peak, we have
found that the correlations functions in the central row behave just like the
one-dimensional regular Ising model and are given by (\ref{corr1gto}).

For $T<T_{\mathrm{c}}(1,m,n)$, the asymptotic behaviors of correlations functions
near the critical point are also two-dimensional regular Ising like, given by (\ref{corr1lt})
for the central rows of strips and (\ref{corr1ltp}) for the centers of string rows. 

In the limit $n\to\infty$, the model is equivalent to independent strips of length $m$,
with its critical temperature at $T=0$. The correlations function at the central row of a strip
is given by (\ref{corr1fn}).

\newpage

\appendix
\section{}
In this appendix we shall compare the approach of the main text with the one of \cite{McCoyPerk}.
It is easily seen that we must identify
\be
(\sigma_j^x,\sigma_j^y,\sigma_j^z)_{\mbox{\small in eq.~(\ref{gamma})}}=
(\sigma_{j-\mathcal{N}-1}^z,-\sigma_{j-\mathcal{N}-1}^y,
\sigma_{j-\mathcal{N}-1}^x)_{\mbox{\small in \cite{McCoyPerk}}},
\label{A1}\ee
comparing (\ref{gamma}) in which $\bar p$ is the horizontal size of the model with (2.7) \cite{McCoyPerk}
where $2\mathcal{N}+1$ is this size, so that for $1\le j\le\bar p=2\mathcal{N}+1$, 
we have\footnote{%
It is easy to change this for the case $\bar p=2\mathcal{N}$ even, but that is not needed here as we
consider the ferromagnetic case in the limit $\bp\to\infty$.}
\be
\gamma_j=2^{-1/2}\Gamma_{j+2\mathcal{N}+1},\quad j=-2\mathcal{N},\cdots,2\mathcal{N}+1.
\label{A2}\ee
As we shall only consider the limit $\bar p,\mathcal{N}\to\infty$ and only consider pair correlations in
rows, we only have to deal with the ``even sector'' and we may modify the boundary conditions to
strict periodicity,
\be
\gamma_{j\pm2\bar p}\equiv\gamma_j,\quad \Gamma_{j\pm2\bar p}\equiv\Gamma_j,
\label{A3}\ee
just as we have already done in the main text and in \cite{McCoyPerk}. This modification,
corresponding to setting $U\equiv1$ in (\ref{V2}), does not affect the final result in the
large-$\bar p$ limit. (Alternatively we could also choose strict antiperiodic boundary conditions
corresponding to $U\equiv-1$, leading to the same final result.)

The setup in \cite{McCoyPerk} has reflection invariance about row 0 and translational invariance in the
horizontal direction, so that the correlation of a spin pair in horizontal row 0 is given by a Toeplitz
determinant. Let us label a given row of horizontal couplings and the row of vertical couplings
directly above it by $r$ (called $\mathrm{m}$ in \cite{McCoyPerk}). The rows labeled by $r$ have
reduced horizontal coupling $\mathrm{H}_r=J'_r/k_{\mathrm{B}}T$ and reduced vertical coupling
$\mathrm{V}_r=J_r/k_{\mathrm{B}}T$.  Reflection invariance requires
\be
\mathrm{H}_{-r}=\mathrm{H}_r\equiv K_{2r},\qquad \mathrm{V}_{-r-1}=\mathrm{V}_r\equiv K_{2r+1},
\qquad K_{-l}=K_l.
\label{A4}\ee
This generalizes the case considered in in the main body of this paper where $J'_r=J'$ or 0 and $J_r=J$,
see figure 1. 
We can write,
\be
\mathrm{T}_l=\exp\Bigg(\sum_{j=-2\mathcal{N}}^{2\mathcal{N}+1}\sum_{k=-2\mathcal{N}}^{2\mathcal{N}+1}
\ri C^{(l)}_{jk}\gamma_j\gamma_k\Bigg),
\label{A5}\ee
where the nonvanishing matrix elements of the $2\times2$ block-cyclic matrices $\mathsf{C}^{l}$ are
\be
C^{(2r)}_{2n-1,2n}=-C^{(2r)}_{2n,2n-1}=\mathrm{H}_r,\qquad
C^{(2r+1)}_{2n,2n+1}=-C^{(2r+1)}_{2n+1,2n}=\mathrm{V}^{\ast}_r,
\label{A6}\ee
setting $2\mathcal{N}+2\equiv-2\mathcal{N}$, compare (2.10)--(2.12) in \cite{McCoyPerk}.
One easily checks that from (\ref{A4}) we have the required
\be
\mathrm{T}_{-l}=\mathrm{T}_l=\mathrm{T}_l^{\;\dagger},
\label{A7}\ee
where we added that $\mathrm{T}_l$ is hermitian, compare (\ref{hV13})--(\ref{hV2a}).

With these notations (2.6) in \cite{McCoyPerk} can be rewritten as\footnote{%
In \cite{McCoyPerk} demanding vertical periodicity $\mathcal{M}+1\equiv-\mathcal{M}$ implies the
vertical size $L=2\mathcal{M}+1$ to be odd. We allow the vertical size $L$ to also be even, i.e.\ vertical
periodicity $\mathcal{M}\equiv-\mathcal{M}$ and $L=2\mathcal{M}$.}
\be
\langle\sigma_{00}\sigma_{0N}\rangle=
\frac{\Tr\bT^{(<)}\sigma^x_0\sigma^x_N\bT^{(>)}}{\Tr\bT^{(<)}\bT^{(>)}}
=\frac{\Tr\sigma^x_0\sigma^x_N\bT^{(0)}}{\Tr\bT^{(0)}},\quad
\bT^{(0)}=\bT^{(>)}\bT^{(<)},
\label{A8}\ee
where
\be
\bT^{(>)}=\mathrm{T}_{0}^{1/2}\Bigg[\prod_{l=1}^{L-1}
\mathrm{T}_{l}\Bigg]\mathrm{T}_{L}^{\;1/2},\quad
\bT^{(<)}=\mathrm{T}_{-L}^{\;1/2}\Bigg[\prod_{l=-L+1}^{-1}
\mathrm{T}_{l}\Bigg]\mathrm{T}_{0}^{1/2}=\bT^{(>)}{}^{\dagger}.
\label{A9}\ee
The equality $\bT^{(<)}=\bT^{(>)}{}^{\dagger}$ expresses the reflection symmetry causing
the Toeplitz determinant to emerge for general layered systems of infinite horizontal size and
translationally invariant in the horizontal direction.

This setup generalizes  the case studied in this paper with $L=p(m+n)$ and vertical period $L_1=m+n$.
We note that $\mathrm{T}_{2r}$, also given in (2.10) of \cite{McCoyPerk}, generalizes $V_2$ in (\ref{V2})
[or the unit matrix $\mathbf{1}$ when $J'_r=0$], whereas $\mathrm{T}_{2r+1}$, also
in (2.11) of \cite{McCoyPerk}, generalizes $V_1$ in (\ref{V1}) and (\ref{V3}), ($V_3=V_1^n$),
omitting the scalar factors
$[2\sinh(2\mathrm{V}_r)]^{\bar p/2}$ that cancel out of the pair correlations. 

For $m$ even, we compare (\ref{A8}) and (\ref{A9}) with (\ref{corr1}) and (\ref{tm}), identifying
the mid-strip row considered as row $r=0$. Now $n$ can be even or odd.
We have $\mathrm{T}_{2r}=V_2$ for $-\half m\le r\le\half m$,
$\mathrm{T}_{2r}=\bf 1$ for $\half m<r<L_1-\half m$, both for $r$ mod $L_1$, whereas
$\mathrm{T}_{2r+1}=V_1$ for all $r$, ($V_3=V_1^n$).
It is easily seen that $\bT^{(0)}={\bT}^p$ with $\bT$ given in (\ref{tm}) and  $\bT^{(0)}$  in (\ref{A8}).

While considering the mid-string row case of (\ref{corr1p}), we have to identify row $\bar n$
as row $r=0$ and now we have to have $n$ even, but $m$ can be even or odd.
Comparing (\ref{corr1p}) and (\ref{htm}) with  (\ref{A8}) and (\ref{A9}), we have
$\mathrm{T}_{2r}=\bf 1$ for $-\half n<r<\half n$, $\mathrm{T}_{2r}=V_2$ for $\half n\le r\le L_1-\half n$, 
both for $r$ mod $L_1$, whereas again $\mathrm{T}_{2r+1}=V_1$ for all $r$, $V_3=V_1^n$.
Now it is easily seen that $\bT^{(0)}={\bT'}^p$ with $\bT'$ given in (\ref{htm}) and  $\bT^{(0)}$  in (\ref{A8}).

Applying the Wick theorem, we find from (\ref{A8})
\be\fl
\langle\sigma_{00}\sigma_{0N}\rangle=
\frac{\Tr\prod_{k=1}^N(2\ri\gamma_{2k-1}\gamma_{2k})\bT^{(0)}}{\Tr\bT^{(0)}}=
\underset{1\le k<l\le2N}{\mathrm{Pf}}\mathrm{G}_{kl},
\qquad\mathrm{G}_{kl}\equiv2\ri\frac{\Tr\gamma_{k}\gamma_{l}\bT^{(0)}}{\Tr\bT^{(0)}},
\label{A10}\ee
compare of (2.16) \cite{McCoyPerk}. With the choice of interactions in the main text 
$\mathrm{G}_{kl}$ equals (\ref{Gp}) or  (\ref{Gpp}) after suitable shifts of the indices.
We calculate $\mathrm{G}_{kl}$ by the method introduced in (4.19) and following text
of \cite{PerkCapel}, as used in \cite{McCoyPerk} starting with
\be
2\ri\delta_{kl}=\mathrm{G}_{kl}+\mathrm{G}_{lk}=
\mathrm{G}_{kl}+2\ri\frac{\Tr({\bT^{(0)}}^{-1}\gamma_{k}\bT^{(0)})\gamma_{l}\bT^{(0)}}{\Tr\bT^{(0)}},
\label{A11}\ee
compare also  (\ref{GpG}) and following text.

In order to calculate this recursively, let us define more generally, (with in the second product the factors
in opposite order as $l$ decreases),
\ba
\bT^{(r)}\equiv\Bigg[\prod_{l=r}^{L-1}\mathrm{T}_{l}\Bigg]\mathrm{T}_{L}
\Bigg[\prod_{l=L-1}^{r}\mathrm{T}_{l}\Bigg],
\quad(0<r<L),\nonumber\\
\bT^{(0)}=\mathrm{T}_{0}^{1/2}\bT^{(1)}\mathrm{T}_{0}^{1/2},\quad
\bT^{(L)}=\mathrm{T}_{L},\quad\bT^{(L+1)}=\mathbf{1},
\label{A12}\ea
so that
\ba
\bT^{(r)}=\mathrm{T}_r' \bT^{(r+1)}\mathrm{T}_r',\nonumber\\
\mathrm{T}_r'\equiv\mathrm{T}_r,\quad\mbox{except}\quad
\mathrm{T}_0'\equiv\mathrm{T}_{0}^{1/2},\quad\mathrm{T}_L'\equiv\mathrm{T}_{L}^{1/2}.
\label{A13}\ea
Also define
\be
\mathrm{G}_{kl}^{(r)}\equiv2\ri\frac{\Tr\gamma_{k}\gamma_{l}\bT^{(r)}}{\Tr\bT^{(r)}},
\quad\mathrm{G}_{kl}^{(L+1)}=\ri\delta_{jk}.
\label{A14}\ee
It is not difficult to show from (\ref{A5}) and (\ref{A6}) that
\be
\mathrm{T}_{l}^{\;-1}\boldsymbol{\gamma}\mathrm{T}_{l}=
\exp(2\ri\mathsf{C}^{(l)})\!\cdot\!\boldsymbol{\gamma},\quad
\mathrm{T}_{l}^{\;-1}\gamma_j\mathrm{T}_{l}=\sum_{k=-2\mathcal{N}}^{2\mathcal{N}+1}
\exp(2\ri\mathsf{C}^{(l)})_{jk}\gamma_k.
\label{A15}\ee
This is nothing but the relation of the spinor and vector representations of the complex
rotation group, advocated by Kaufman in her solution of the 2D Ising model \cite{Kaufman} and
given in equivalent notation in (\ref{cVi}).
Now defining, in analogy with (\ref{A12}) and (\ref{A13}),
\be
\check\mathsf{T}^{(r)}=\Bigg[\prod_{l=r}^{L}\check\mathsf{T}_{l}'\Bigg]\Bigg[\prod_{l=L}^{r}\check\mathsf{T}_{l}'\Bigg],\quad
\check\mathsf{T}_{l}\equiv\exp(2\ri\mathsf{C}^{(l)}),
\label{A16}\ee
where the check on these matrices indicate that they are $2\bp\times2\bp$, not the $2^{\bp}\times2^{\bp}$
$\bT^{(r)}$ and $\mathrm{T}_{l}$, and
where the prime means that for $l=0$ and $l=L$, matrix $\mathsf{C}^{(l)}$ must be replaced by
$\frac12\mathsf{C}^{(l)}$, as implied by (\ref{A13}). From (\ref{A11})  and (\ref{A14}) we then find in matrix notation
\be
2\ri\mathbf{1}=\mathsf{G}^{(r)}+\check\mathsf{T}^{(r)}\!\cdot\!\mathsf{G}^{(r)},\quad
\mathsf{G}^{(r)}=2\ri(\mathbf{1}+\check\mathsf{T}^{(r)})^{-1},\quad
(0\le r\le L+1),
\label{A17}\ee
compare (2.20) in \cite{McCoyPerk}, (4.35) and (4.38) in \cite{PerkCapel} and (38) in the main text.

As the $\mathsf{C}^{(l)}$ are $2\times2$ block-cyclic, so are all matrices in  (\ref{A16})
and  (\ref{A17}). Let us apply the discrete block-Fourier transform on all such matrices
$\mathsf{X}$,
\be
\hat\mathsf{X}\equiv\hat\mathsf{X}(\theta)\equiv\frac{1}{2\mathcal{N}+1}
\sum_{j=-\mathcal{N}}^{\mathcal{N}}\sum_{k=-\mathcal{N}}^{\mathcal{N}}
\re^{\ri\theta(k-j)}
\left(\begin{array}{cc}
X_{2j-1,2k-1}&X_{2j-1,2k}\\
X_{2j,2k-1}&X_{2j,2k}
\end{array}\right),
\label{A18}\ee
for $\theta\equiv\theta_s=2\pi s/(2\mathcal{N}+1)$, $s=-\mathcal{N},\ldots,\mathcal{N}$.
The hat on the matrix indicates that is a $2\times2$ block of the block-diagonal Fourier transform.
In the limit $\mathcal{N}\to\infty$ the inverse Fourier transform is
\be
\left(\begin{array}{cc}
X_{2j-1,2k-1}&X_{2j-1,2k}\\
X_{2j,2k-1}&X_{2j,2k}
\end{array}\right)
=\frac{1}{2\pi}\int_{-\pi}^{\pi}\rd\theta\,
\re^{-\ri\theta(k-j)}\hat\mathsf{X}(\theta).
\label{A19}\ee
Using (\ref{A6}) and (\ref{A18}) one easily verifies
\be
\hat\mathsf{C}^{(2r)}=\left(\begin{array}{cc}
0&\mathrm{H}_r\\
-\mathrm{H}_r&0
\end{array}\right),\quad
\hat\mathsf{C}^{(2r+1)}=\left(\begin{array}{cc}
0&-\mathrm{V}_r^{\ast}\,\re^{-\ri\theta}\\
\mathrm{V}_r^{\ast}\,\re^{\ri\theta}&0
\end{array}\right),
\label{A20}\ee
so that the Fourier transforms of $\check\mathsf{T}_{l}$ become the SU(1,1) matrices
\ba\fl
\hat\mathsf{T}_{2r}=\exp(-2\mathrm{H}_r\sigma^y),\quad
\hat\mathsf{T}_{2r+1}=\exp(-2\mathrm{V}_r^{\ast}\sigma^{\theta}),\quad
\sigma^{\theta}\equiv\sigma^x\sin\theta-\sigma^y\cos\theta,\nonumber\\
\fl
\hat\mathsf{T}_{2r}=\left(\begin{array}{cc}
\cosh(2\mathrm{H}_r)&\ri\sinh(2\mathrm{H}_r)\\
-\ri\sinh(2\mathrm{H}_r)&\cosh(2\mathrm{H}_r)
\end{array}\right),\quad
\hat\mathsf{T}_{2r}^{1/2}=\left(\begin{array}{cc}
\cosh(\mathrm{H}_r)&\ri\sinh(\mathrm{H}_r)\\
-\ri\sinh(\mathrm{H}_r)&\cosh(\mathrm{H}_r)
\end{array}\right),\nonumber\\
\fl
\hat\mathsf{T}_{2r+1}=\left(\begin{array}{cc}
\cosh(2\mathrm{V}_r^{\ast})&-\ri\re^{-\ri\theta}\sinh(2\mathrm{V}_r^{\ast})\\
\ri\re^{\ri\theta}\sinh(2\mathrm{V}_r^{\ast})&\cosh(2\mathrm{V}_r^{\ast})
\end{array}\right),\qquad
\hat\mathsf{T}_{l}^{\phantom{y}}=\hat\mathsf{T}_{l}^{\dagger},\quad
\hat\mathsf{T}_{l}'=\hat\mathsf{T}_{l}'{}^{\dagger}.
\label{A21}\ea
Therefore, any product $\hat\mathsf{D}$ of $\hat\mathsf{T}_{l}$'s or their inverses is of the form
(2.25) in \cite{McCoyPerk},
\be
\hat\mathsf{D}=\left(\begin{array}{cc}
x&\ri y\\
-\ri\overline{y}&\overline{x}
\end{array}\right)=
\exp(a_x\sigma^x+a_y\sigma^y+a_z\ri\sigma^z),\quad
\det \hat\mathsf{D}=1,
\label{A22}\ee
where overlining denotes complex conjugate, $a_x,a_y,a_z$ are three real parameters, and
$\sigma^x,\sigma^y,\ri\sigma^z$ are generators of SU(1,1) related to hyperbolic geometry
already present in the papers of Onsager and Kaufman \cite{Kaufman,Onsager}.
The hermitian conjugate $\hat\mathsf{D}^{\dagger}$
is the product with the factors in opposite order (not equal $\hat\mathsf{D}$ in general), for example,
\be
\hat\mathsf{D}=\hat\mathsf{T}_{k}'\hat\mathsf{T}_{k+1}'\cdots\hat\mathsf{T}_{l-1}'\hat\mathsf{T}_{l}'
\quad\Longleftrightarrow\quad
\hat\mathsf{D}^{\dagger}=\hat\mathsf{T}_{l}'\hat\mathsf{T}_{l-1}'\cdots\hat\mathsf{T}_{k+1}'\hat\mathsf{T}_{k}'.
\label{A23}\ee

From the Fourier transforms of (\ref{A13})--(\ref{A17}) we find
\be\fl
\hat\mathsf{G}^{(r)}=2\ri(\mathbf{1}_2+\hat\mathsf{T}^{(r)})^{-1},\quad
\hat\mathsf{T}^{(r)}=\hat\mathsf{T}_r'\hat\mathsf{T}^{(r+1)}\hat\mathsf{T}_r',\quad
\hat\mathsf{G}^{(L+1)}=\ri\mathbf{1}_2=
\Bigg(\begin{array}{cc}
\ri&0\\
0&\ri
\end{array}\Bigg).
\label{A24}\ee
Now using (\ref{A23}) with $k=r$ and $l=s-1$ in (\ref{A24}), we go though the steps
\ba\fl
\hat\mathsf{T}^{(r)}=\hat\mathsf{D}\hat\mathsf{T}^{(s)}\hat\mathsf{D}^{\dagger},\quad
\hat\mathsf{G}^{(r)}-2\ri\mathbf{1}_2=-\hat\mathsf{G}^{(r)}\hat\mathsf{T}^{(r)}=
-\hat\mathsf{G}^{(r)}\hat\mathsf{D}\hat\mathsf{T}^{(s)}\hat\mathsf{D}^{\dagger},\nonumber\\
\fl
(\hat\mathsf{G}^{(r)}-2\ri\mathbf{1}_2)\hat\mathsf{D}^{\dagger}{}^{-1}\hat\mathsf{G}^{(s)}
=-\hat\mathsf{G}^{(r)}\hat\mathsf{D}\hat\mathsf{T}^{(s)}\hat\mathsf{G}^{(s)},\quad
\hat\mathsf{G}^{(s)}-2\ri\mathbf{1}_2=-\hat\mathsf{T}^{(s)}\hat\mathsf{G}^{(s)},
\label{A25}\ea
to arrive at
\be
(\hat\mathsf{G}^{(r)}-2\ri\mathbf{1}_2)\hat\mathsf{D}^{\dagger}{}^{-1}\hat\mathsf{G}^{(s)}
=\hat\mathsf{G}^{(r)}\hat\mathsf{D}(\hat\mathsf{G}^{(s)}-2\ri\mathbf{1}_2).
\label{A26}\ee
Assuming $\hat\mathsf{G}^{(s)}$ has the form
\be
\hat\mathsf{G}^{(s)}=
\Bigg(\begin{array}{cc}
\ri&\mathrm{K}_s\\
-\overline{\mathrm{K}}_s&\ri
\end{array}\Bigg),
\label{A27}\ee
consistent with $\hat\mathsf{G}^{(L+1)}=\ri\mathbf{1}_2$ in (\ref{A24}),
and let $\hat\mathsf{D}$ be given by (\ref{A22}), then we can solve $\hat\mathsf{G}^{(r)}$
from the four linear equations for its four entries given by (\ref{A26}). The solution is
\be
\hat\mathsf{G}^{(r)}=
\Bigg(\begin{array}{cc}
\ri&\mathrm{K}_r\\
-\overline{\mathrm{K}}_r&\ri
\end{array}\Bigg),\qquad
\mathrm{K}_r=\frac{y+x\mathrm{K}_s}{\overline{x}+\overline{y}\mathrm{K}_s},
\label{A28}\ee
compare (2.32) in \cite{McCoyPerk}. Since (\ref{A28}) implies
$\mathsf{G}_{2j,2k}^{(r)}=\mathsf{G}_{2j-1,2k-1}^{(r)}=\ri\delta_{jk}$ for all $r$, the Pfaffian (\ref{A10})
for the pair correlation function reduces to the Toeplitz determinant
\be\fl
\langle\sigma_{00}\sigma_{0N}\rangle=\det_{1\le j,k\le N}\mathsf{A},\quad
A_{jk}=G_{2j-1,2k}=a_{j-k},\quad a_n=\frac{1}{2\pi}\int_{-\pi}^{\pi}\rd\theta\,
\re^{-\ri n\theta}\mathrm{K}(\re^{\ri\theta}),
\label{A29}\ee
Thus we obtain a recurrence relation for the
generating function $\mathrm{K}(\re^{\ri\theta})\equiv\mathrm{K}_0$.

This result applies to all layered systems with reduced couplings as specified in 
(\ref{A4}) and infinite horizontal size. If  $L=2\mathcal{M}=0$, we have just one
horizontal chain and (\ref{A9}) degenerates to $\bT^{(>)}=\bT^{(<)}=\mathrm{T}_{0}^{1/2}$;
then $\mathrm{K}_1=0$ and $\mathrm{K}_0=\tanh(\mathrm{H}_0)$, so that
$\mathsf{A}=\mathbf{1}_N\tanh(\mathrm{H}_0)$ is a constant diagonal matrix, reproducing the
well-known $\langle\sigma_{00}\sigma_{0N}\rangle=\tanh^N(\beta J_0)$.

If the vertical size is finite, generating function $\mathrm{K}$ is a rational function of $\re^{\ri\theta}$
and $\langle\sigma_{00}\sigma_{0N}\rangle$ has exponential decay in $N$.
It is easy to generate many explicit examples of such $\mathrm{K}$ with the formalisms in
this paper with either open or closed boundary at $L$. In the open cases we take
$L$ odd and $\hat\mathsf{T}_L$ given in (\ref{A21}) with corresponding $\mathrm{V}_r=0$,
$\mathrm{V}_r^{\ast}=\infty$, compare also section 3.3 and \cite[eqs.~(24)--(26)]{HJPih}
and (\ref{A49}) below.

If the vertical size $L$ is also infinite with couplings periodically repeated with period $L_1$,
we can show that we end up with $\hat\mathsf{G}^{(0)}=\hat\mathsf{G}^{(L_1)}$.
Then, if $\overline{x}=x$ in (\ref{A22}) as is true in cases of interest, we have from (\ref{A28})
\be
\mathrm{K}=\frac{y+x\mathrm{K}}{\overline{x}+\overline{y}\mathrm{K}}=
\frac{y+x\mathrm{K}}{x+\overline{y}\mathrm{K}},\quad
\mathrm{K}=\mathrm{K}_0=\pm\sqrt{y/\overline{y}}.
\label{A30}\ee
In order to determine the sign, we set $\theta=\pm\pi$ and assume the Ising
couplings $\mathrm{H}_r,\mathrm{V}_r$ to be non-negative and not all zero.
(Negative signs can be moved to boundary conditions by gauge transformation.)
Then all matrices in (\ref{A21}) have the form (\ref{A22}) with $x>0$, $y\ge0$ real.
Starting with $\mathrm{K}^{(L+1)}=0$ implied by (\ref{A24}), (\ref{A28}) implies 
all $\mathrm{K}^{(r)}\ge0$ and $>0$ from some point on. So we need the
sign in (\ref{A30}) that makes $\mathrm{K}>0$ for $\theta=\pm\pi$.

We have kept manifest reflection symmetry in the above. Once this is broken one will have
to deal with 2$\times$2 generating functions of 2$\times$2 block-Toeplitz determinants.
To generalize the formalism, we have to allow the $\hat\mathsf{D}$ and 
$\hat\mathsf{D}^{\dagger}$ in (\ref{A25}) and (\ref{A26}) to become unrelated products of
$\hat\mathsf{T}_{l}$'s. Then one can derive
\be\fl
\hat\mathsf{T}^{(r)}=\hat\mathsf{D}_1\hat\mathsf{T}^{(s)}\hat\mathsf{D}_2^{\dagger}
\;\Longrightarrow\;
(\hat\mathsf{G}^{(r)}-2\ri\mathbf{1}_2)(\hat\mathsf{D}_2^{\dagger})^{-1}\hat\mathsf{G}^{(s)}
=\hat\mathsf{G}^{(r)}\hat\mathsf{D}_1(\hat\mathsf{G}^{(s)}-2\ri\mathbf{1}_2).
\label{A31}\ee
Here $\hat\mathsf{D}_1$ or $\hat\mathsf{D}_2$ may also be $\mathbf{1}_2$. Typically
we have to apply this $\bar L$ times to calculate $\hat\mathsf{G}=\hat\mathsf{G}^{(0)}$,
starting with $\hat\mathsf{G}^{(\bar L)}=\ri\mathbf{1}_2$ and using $\hat\mathsf{T}^{(i-1)}=
\hat\mathsf{D}_1^{(i)}\hat\mathsf{T}^{(i)}\hat\mathsf{D}_2^{(i)}{\vphantom{h^h}}^{\dagger}$,
$i=\bar L,\bar L-1,\cdots,2,1$, in (\ref{A31}) to solve $\hat\mathsf{G}^{(i)}$ recursively.
If $\hat\mathsf{D}_2^{(i)}=\hat\mathsf{D}_1^{(i)}$ for all $i$, the solutions obey (\ref{A28}).

In order to determine the most general form of $\hat\mathsf{G}=\hat\mathsf{G}^{(0)}$,
we set $r=0$, $s=\bar L=1$, $\hat\mathsf{D}_1=\hat\mathsf{T}^{(0)}$
(corresponding to the full Fourier-transformed transfer matrix representative),
$\hat\mathsf{D}_2=\mathbf{1}_2$ and $\hat\mathsf{G}^{(\bar L)}=\ri\mathbf{1}_2$
 in (\ref{A26}). As $\hat\mathsf{D}$ is a
product of matrices in (\ref{A21}), it is of the form (\ref{A22}) with
suitable $x$ and $y$ being Laurent polynomials in $\re^{\ri\theta}$,
$x\bar x-y\bar y=1$. We then can solve the four linear equations (\ref{A31}) and find
\be
\hat\mathsf{G}=
\left(\begin{array}{cc}
\frac{2\ri(\overline{x}+1)}{x+\overline{x}+2}&\frac{2y}{x+\overline{x}+2}\\
\frac{-2\overline{y}}{x+\overline{x}+2}&\frac{2\ri(x+1)}{x+\overline{x}+2}
\end{array}\right).
\label{A32}\ee
We note that $\hat\mathsf{G}_{22}=-\overline{\hat\mathsf{G}}_{11}$,
$\hat\mathsf{G}_{21}=-\overline{\hat\mathsf{G}}_{12}$,
$\hat\mathsf{G}_{11}+\hat\mathsf{G}_{22}=2\ri$, consistent with the first equality in
(\ref{A11}).\footnote{%
One may check the effect of block-Fourier transform after applying spatial reflection
and complex conjugation to $\mathsf{G}$. Also, $\mathsf{G}$ replaced by
$2\ri\mathbf{1}$---the diagonal matrix with each diagonal $2\times2$ block replaced by the
trace of the corresponding block in $\mathsf{G}$ times $\mathbf{1}_2$---is invariant
under block-Fourier transform.}
This is the same structure as found long ago in a special case  \cite{AuYangMcCoy2},
where the horizontal couplings are left uniform and the structure is less understood.
The above choice $\hat\mathsf{D}_2=\mathbf{1}_2$ is not the only useful one.
Making suitable choices of $r$, $s$, $\hat\mathsf{D}_1$ and $\hat\mathsf{D}_2$ in
(\ref{A31}), we can derive several new results for 2$\times$2 generating functions
for pair correlations within a horizontal row of such layered models by
solving the subsequent sets of four linear equations for the subsequent $\hat\mathsf{G}^{(r)}$.

When $x(\theta)$ is real, $\overline{x}\equiv x$, (\ref{A32}) reduces to the form (\ref{A27}) with
$\mathrm{K}=\mathrm{K}_0=y/(x+1)$, which in the limit of vertical size $L\to\infty$,
$x,y\to\infty$, $y\overline{y}/x^2\to1$, reduces to $\mathrm{K}=\sqrt{y/\overline{y}}$, of the
form (\ref{A30}) with different $y$. This is not surprising: Consider the special case
with some $\hat\mathsf{D}$, with $x=\cosh(\chi)$, $y=\re^{\ri\psi}\sinh(\chi)$, and
$\chi$ and $\psi$ real, repeated $p$ times; then $\hat\mathsf{D}_1=\hat\mathsf{D}^p$ has
$x=\cosh(p\chi)$, $y=\re^{\ri\psi}\sinh(p\chi)$, so that the ratio $y/\overline{y}$ is
unchanged, even though $x$ and $y$ blow up as $p\to\infty$. This parametrization
makes it clear that (\ref{A32}) is equivalent to (\ref{bGmatrix0}) replacing the $x$ there
by $-\re^{\ri\psi}$ and $\Lambda$ by $\re^{\chi}$, and (\ref{bGmatrix}) follows
similarly as $p\to\infty$.

The above formalism is very flexible and has been applied first
\cite{McCoyPerk,McCoyPerk1} to the special line in the Bariev problem \cite{Bariev},
and later to go off that line to calculate energy-density pair correlations
\cite{KoAYP}. The methods can also be applied to calculate 2$\times$2 block-Toeplitz
determinants for the layered model of \cite{HAYFisher,HAY}. However, reflection symmetry
in those models can be restored by the minor modification of changing the horizontal
couplings in the boundaries between layers with different couplings to the average
of those on both sides, or any other reasonable equal value for all boundary rows.
Then the pair correlations in the centers of layers are given by scalar Toeplitz
determinants. The models in this paper and the two preceding ones
\cite{HJPih,HJPihd} are such models.

From now on we consider an Ising model with $p$ infinite horizontal strips strips of width
$m_1$ and reduced couplings $\mathrm{H}_1$ and  $\mathrm{V}_1$ alternating with $p$
strips of width $m_2$ and couplings $\mathrm{H}_2$ and  $\mathrm{V}_2$. On the
horizontal line between two such strips we choose horizontal coupling
$\mathrm{H}_3=\frac12(\mathrm{H}_1+\mathrm{H}_2)+\Delta\mathrm{H}$. We shall also
consider more general insertions.

As usually done for the uniform case, we work out the symmetrized single-row transfer matrices
\be\fl
\hat\mathsf{D}_j\equiv
\exp(-\mathrm{H}_j\sigma^y)\exp(-2\mathrm{V}_j^{\ast}\sigma^{\theta})\exp(-\mathrm{H}_j\sigma^y)
=\left(\begin{array}{cc}
x_j^{(1)}&\ri y_j^{(1)}\\
-\ri\overline{y}_j^{(1)}&\overline{x}_j^{(1)}
\end{array}\right),\quad j=1,2.
\label{A33}\ee
Using (\ref{A21}) we find
\ba\fl
x_j^{(1)}=\cosh(2\mathrm{H}_j)\cosh(2\mathrm{V}_j^{\ast})-
\sinh(2\mathrm{H}_j)\sinh(2\mathrm{V}_j^{\ast})\cos\theta,
\nonumber\\
\fl
y_j^{(1)}=\sinh(2\mathrm{H}_j)\cosh(2\mathrm{V}_j^{\ast})-
\cosh(2\mathrm{H}_j)\sinh(2\mathrm{V}_j^{\ast})\cos\theta
+\ri\sinh(2\mathrm{V}_j^{\ast})\sin\theta,
\nonumber\\
\fl
x_j^{(1)}=\cosh\gamma_j,\quad
y_j^{(1)}=\sinh\gamma_j(\cos\delta_j^{\ast}+\ri\sin\delta_j^{\ast})=
\re^{\ri\delta_j^{\ast}}\sinh\gamma_j,
\label{A34}\ea
compare (2.36) in \cite{McCoyPerk}.
Eq.~(\ref{A34}) with hyperbolic angle $\gamma_j$ and angle $\delta_j^{\ast}$ first appeared
in eq.~(89) of \cite{Onsager} with minor differences of notation.

For $m$ such rows we get $(\hat\mathsf{D}_j)^m$ and $(\hat\mathsf{D}_j)^{m/2}$
expressed by similar formulae, just replacing $\gamma_j$ by $m\gamma_j$  or
$\frac12m\gamma_j$ according to the SU(1,1) group structure,
\ba\fl
x_j^{(m)}=\cosh(m\gamma_j)=\mathrm{T}_m(\cosh\gamma_j)=\mathrm{T}_m(x_j^{(1)}),\nonumber\\
\fl
y_j^{(m)}=\re^{\ri\delta_j^{\ast}}\sinh(m\gamma_j)=
(\re^{\ri\delta_j^{\ast}}\sinh\gamma_j)\mathrm{U}_{m-1}(\cosh\gamma_j)=
y_j^{(1)}\mathrm{U}_{m-1}(x_j^{(1)}),
\label{A35}\ea
where $\mathrm{T}_j(x)$ and  $\mathrm{U}_j(x)$ are the Chebyshev polynomials of the
first and second kind. Thus $x_j^{(m)}$ and $y_j^{(m)}$ are explicitly expressed as Laurent
polynomials in $\re^{\ri\theta}$. When $m$ is odd, multiplying by
$2x_j^{(1/2)}=2\cosh(\gamma_j/2)$  and working out the products we get
\ba\fl
x_j^{(m/2)}=\left(\mathrm{T}_{(m+1)/2}(x_j^{(1)})+\mathrm{T}_{(m-1)/2}(x_j^{(1)})\right)
/\left(2\cosh(\gamma_j/2)\right),\nonumber\\
\fl
y_j^{(m/2)}=
y_j^{(1)}\left(\mathrm{U}_{(m-1)/2}(x_j^{(1)})+\mathrm{U}_{(m-3)/2}(x_j^{(1)})\right)
/\left(2\cosh(\gamma_j/2)\right),
\label{A36}\ea
identifying $\mathrm{U}_{-1}(x)\equiv0$. The denominators $2x_j^{(1/2)}$
cancel out in later calculations.

We remark that we need to work with $(\hat\mathsf{D}_j)^{m_j/2}$, $(j=1,2)$, in order to keep
reflection symmetry manifestly. Of course, $m_j$ has to be even if we calculate the pair correlation
in the center row of a layer of type $j$. Between the layers of type 1 and 2 we have alternatingly 
$\hat\mathsf{D}_3$ and $\hat\mathsf{D}_3{}^{\!\dagger}$, corresponding to the product of some
transfer matrices and the product in the opposite order. Then, depending on $j=1$ or 2 for the layer
in which the correlations are calculated, the full Fourier-transformed transfer matrix becomes
\be
\hat\mathsf{T}=\left(\hat\mathsf{T}_{j\leftarrow j}\right)^p,\quad
\hat\mathsf{T}_{1\leftarrow1}=\hat\mathsf{T}_{1\leftarrow2}\hat\mathsf{T}_{2\leftarrow1},\quad
\hat\mathsf{T}_{2\leftarrow2}=\hat\mathsf{T}_{2\leftarrow1}\hat\mathsf{T}_{1\leftarrow2},
\label{A37}\ee
where
\be
\hat\mathsf{T}_{1\leftarrow2}=(\hat\mathsf{D}_1)^{m_1/2}\,
\hat\mathsf{D}_3\,(\hat\mathsf{D}_2)^{m_2/2},\quad
\hat\mathsf{T}_{2\leftarrow1}=(\hat\mathsf{D}_2)^{m_2/2}\,
\hat\mathsf{D}_3{}^{\!\dagger}\,(\hat\mathsf{D}_1)^{m_1/2},
\label{A38}\ee
$p\to\infty$. Using the second member of (\ref{A28}) three times, and noting that
$\hat\mathsf{D}_3{}^{\!\dagger}$ has $x_3$ and $\overline{x}_3$  interchanged, we find the actions
\be
\hat\mathsf{T}_{1\leftarrow2}:\;\mathrm{K}\to\frac{Q+P\mathrm{K}}{\overline{P}+\overline{Q}\mathrm{K}},\quad
\hat\mathsf{T}_{2\leftarrow1}:\;\mathrm{K}\to\frac{Q+\overline{P}\mathrm{K}}{P+\overline{Q}\mathrm{K}},
\label{A39}\ee
with
\ba\fl
P=x_1^{(m_1/2)}x_2^{(m_2/2)}x_3+x_1^{(m_1/2)}\overline{y}_2^{(m_2/2)}y_3+
y_1^{(m_1/2)}x_2^{(m_2/2)}\overline{y}_3+
y_1^{(m_1/2)}\overline{y}_2^{(m_2/2)}\overline{x}_3,\nonumber\\
\fl
Q=x_1^{(m_1/2)}x_2^{(m_2/2)}y_3+x_1^{(m_1/2)}y_2^{(m_2/2)}x_3+
y_1^{(m_1/2)}x_2^{(m_2/2)}\overline{x}_3+y_1^{(m_1/2)}y_2^{(m_2/2)}\overline{y}_3,
\label{A40}\ea
which relate by $y_2\leftrightarrow\overline{y}_2$, $x_3\leftrightarrow y_3$.
Next, from (\ref{A37}) and (\ref{A28}) we find
\ba
\hat\mathsf{T}_{1\leftarrow1}:\;\mathrm{K}\to
\frac{2PQ+(P\overline{P}+Q\overline{Q})\mathrm{K}}{(P\overline{P}+Q\overline{Q})+2\overline{P}\,\overline{Q}\mathrm{K}},
\nonumber\\
\hat\mathsf{T}_{2\leftarrow2}:\;\mathrm{K}\to
\frac{2\overline{P}Q+(P\overline{P}+Q\overline{Q})\mathrm{K}}{(P\overline{P}+Q\overline{Q})+2P\overline{Q}\mathrm{K}},
\label{A41}\ea
and from (\ref{A30}) we then find the generating functions in the limit $p\to\infty$
\be
\mathrm{K}_1=\sqrt{\frac{PQ}{\overline{P}\,\overline{Q}}},\quad
\mathrm{K}_2=\sqrt{\frac{\overline{P}Q}{P\overline{Q}}},
\label{A42}\ee
for the center rows of layers of type 1 and 2. From (\ref{A39})
 one can check that these also satisfy the required actions
 \be
\hat\mathsf{T}_{1\leftarrow2}:\;\mathrm{K}_2\to\mathrm{K}_1,\quad
\hat\mathsf{T}_{2\leftarrow1}:\;\mathrm{K}_1\to\mathrm{K}_2.
\label{A43}\ee

If we now set $m_1=m=2j$, $m_2=n$, $\mathrm{H}_1=2\Delta\mathrm{H}=J'/k_{\mathrm{B}}T$,
$\mathrm{H}_2=0$, $\mathrm{V}_1=\mathrm{V}_2=J/k_{\mathrm{B}}T$, $\mathrm{V}_3{}^{\!\ast}=0$,
we reproduce (\ref{Phi}) and  (\ref{Phip}) in the main text. Indeed, we must identify
\be\fl
\alpha=\re^{\gamma_1},\quad \Omega^{1/2}=\re^{\delta_1^{\ast}},
\quad z=\tanh(\mathrm{V}_2)=\re^{-2\mathrm{V}_2^{\ast}},
\quad \re^{\delta_2^{\ast}}=-\re^{-\ri\theta},
\quad z'=\tanh(\mathrm{H}_3),
\label{A44}\ee
so that (\ref{A34}) gives
\ba
x_1=\frac{\alpha^j+\alpha^{-j}}{2},\quad y_1=\Omega^{1/2}\,\frac{\alpha^j-\alpha^{-j}}{2},
\quad \overline{y}_1=\Omega^{-1/2}\,\frac{\alpha^j-\alpha^{-j}}{2},\nonumber\\
x_2=\frac{1+z^n}{2z^{n/2}},\quad y_2=-\re^{-\ri\theta}\,\frac{1-z^n}{2z^{n/2}},
\quad \overline{y}_2=-\re^{\ri\theta}\,\frac{1-z^n}{2z^{n/2}},\nonumber\\
x_3=\overline{x}_3=\frac{1}{\sqrt{1-z'{}^2}},\quad y_3=\overline{y}_3=\frac{z'}{\sqrt{1-z'{}^2}}.
\label{A45}\ea
One then easily verifies, comparing (\ref{A39}) with (\ref{Phi}) and  (\ref{Phip}),
 \be
 \frac{A}{\overline{P}}=\frac{B}{\overline{Q}}=\frac{\overline{A}}{P}=\frac{\overline{B}}{Q}
 =4z^{n/2}\sqrt{1-z'{}^2}.
\label{A46}\ee

We end this appendix with a few remarks on the case with uniform interactions,
infinite horizontal size, but finite vertical size $L$. For the infinite cylinder with
circumference $L$, we can use $\hat{\mathsf{D}}$ given in (\ref{A33})--(\ref{A36})
leaving out the $j$-subscripts. When $L$ is odd, we can factor the boundary
$\hat{\mathsf{D}}$ as a product of its square roots, which is just replacing $\gamma$
by $\half\gamma$ because of the SU(1,1) group structure. We can now apply
(\ref{A28}) with $r=0$, $s=L+1$, $x=\bar x=x^{(L/2)}$, $y=y^{(L/2)}$, or
alternatively (\ref{A32}) with $x=\bar x=x^{(L)}$, $y=y^{(L)}$. The result,
valid for all $L$, is
\be
\mathrm{K}(\re^{\ri\theta})=\frac{y^{(L/2)}}{x^{(L/2)}}
=\frac{\re^{\ri\delta^{\ast}}\sinh(L\gamma/2)}{\cosh(L\gamma/2)}
=\frac{y^{(L)}}{x^{(L)}+1}
=\frac{\re^{\ri\delta^{\ast}}\sinh(L\gamma)}{\cosh(L\gamma)+1}.
\label{A47}\ee
The case $L=1$ is extremely simple, as using (\ref{A34}), (\ref{A29}) and the
residue theorem gives the expected Ising chain result,
\ba
\mathrm{K}(\re^{\ri\theta})=\frac{z\re^{\ri\theta}-z^{\ast}}{\re^{\ri\theta}-zz^{\ast}},\quad
z=\tanh(\mathrm{H}),\quad z^{\ast}=\tanh(\mathrm{V}^{\ast}),
\quad a_0=z,\nonumber\\
a_n=(zz^{\ast})^n(z-z^{-1}),\quad a_{-n}=0,\quad (n>0),
\quad \langle\sigma_{00}\sigma_{0N}\rangle=z^{|N|}.
\ea
For $L>1$, $a_{-n}\ne0$, as we found that $\mathrm{K}(w)$, $w=\re^{\ri\theta}$,
typically has $\lceil\half L\rceil$ poles $0<w_j<1$ and $\lfloor\half L\rfloor$ poles $w_j>1$.
Values $w_j$ can be worked out easily to high precision numerically leading to explicit
formulae for $a_n$ of the form $c_0\delta_{n0}+\sum\epsilon_j c_j w_j^n$,
($\epsilon_j=1$ or 0, so that for $n\ge0$ only $w_j<1$ contribute and for $n<0$
only $w_j>1$).
 
The generating function for the middle row of an open strip with free boundaries
and $L$ even can be treated by coupling the two sides with zero coupling. Thus it
becomes the cylinder case with circumference $L+1$, $\mathrm{V}_{\pm L/2}=0$,
$\mathrm{V}^{\ast}_{\pm L/2}=\infty$, compare (24)--(26) in \cite{HJPih}. We can
calculate the generating function by $\bar L=3$ applications of (\ref{A28}). Starting with
$\hat\mathsf{G}^{(3)}=\ri\mathbf{1}_2$, $\mathrm{K}_3=0$ we apply
$\hat\mathsf{D}^{(3)}=\hat\mathsf{T}_{2r+1}$ in (\ref{A21}) with $\mathrm{V}_r^*\to\infty$,
or $y/x\to-\re^{-\ri\theta}$ by comparing with (\ref{A22}). Thus we find
$\mathrm{K}_2=-\re^{-\ri\theta}$ as in (\ref{Phiz0p}). Next we apply
$\hat\mathsf{D}^{(2)}=\hat\mathsf{T}_{2r}$ in (\ref{A21}) with
$\mathrm{H}_r=\half\mathrm{H}$, or $x=\cosh(\mathrm{H})$,
$y=\sinh(\mathrm{H})=z'x$, resulting in
$\mathrm{K}_1=(z'-\re^{-\ri\theta})/(1-z'\re^{-\ri\theta})$. Finally
we apply $\hat\mathsf{D}^{(1)}$ with $x=x^{(L/2)}$,
$y=y^{(L/2)}$, as given in (\ref{A35}) dropping the subscripts $j$ and identifying
$\re^{\ri\delta^*}=\Omega^{1/2}=\overline{\Omega^{-1/2}}$, and arrive at
\be
\mathrm{K}=\frac{(z'-\re^{-\ri\theta})+\Omega^{1/2}\tanh(\half L\gamma)(1-z'\re^{-\ri\theta})}
{(1-z'\re^{-\ri\theta})+\Omega^{-1/2}\tanh(\half L\gamma)(z'-\re^{-\ri\theta})},
\label{A49}\ee
reproducing (\ref{Phiz0}) identifying $\alpha=\re^{\gamma}$, $j=\half L$
and becoming $\Omega^{1/2}$ when $L\to\infty$. Here and in (\ref{A50}) below
$\Omega^{1/2}\tanh(\half L\gamma)$ is the result for the cylindrical case (\ref{A47}).

For the middle row of an open strip of even width $L$ and with fixed boundary
values all equal, we identify the boundary spins in the same column and connect
them with infinite coupling. Now the model is the cylinder case with the same
size $L$ and $\mathrm{H}_{\pm L/2}=\infty$, forcing all these spins to be equal.
Now we apply (\ref{A28}) $\bar L=2$ times. Starting with $\mathrm{K}_2=0$
and applying $\hat\mathsf{D}^{(2)}=\hat\mathsf{T}_{2r}$ 
with $\mathrm{H}\to\infty$, $y/x\to1$ we find $\mathrm{K}_1=1$ from (\ref{A28}).
Repeating the application of $\hat\mathsf{D}^{(1)}$ as in the previous paragraph,
we find
\be
\mathrm{K}=\frac{1+\Omega^{1/2}\tanh(\half L\gamma)}
{1+\Omega^{-1/2}\tanh(\half L\gamma)},
\label{A50}\ee
which in the limit $L\to\infty$ becomes $\Omega^{1/2}$ as it should.
The pair correlation does not depend on the sign of these spins, as it is invariant
under flipping the signs of all spins. Therefore, this case also gives the result
for a strip of width $L-2$ with a constant boundary field. The special case
$L=2$ is also the case of the Ising chain in a field $h=2J$.

Correlations in other rows than the middle one, or with asymmetric boundary
conditions, generally can be expressed as block-Toeplitz determinants whose
2$\times$2 generating functions can also be studied with the methods in this
appendix. We may even move from row to row as needed in \cite{KoAYP}
by applying (\ref{A31}) with
$\hat\mathsf{D}^{(2)}=\hat\mathsf{D}^{(1)}{\vphantom{h^h}}^{-1}$.
 

\end{document}